\documentclass[pra,aps,groupedaddress,floatfix,showpacs,showkeys]{revtex4-1}
\usepackage{graphicx}
\usepackage{amsmath}
\usepackage{lipsum}
\usepackage{multirow}
\usepackage{footnote}

\newcommand{\ket}[1]{\mathop{| #1 \rangle}\nolimits}
\newcommand{\matelem}[3]{\mathop{\langle #1 | #2 | #3\rangle}\nolimits}

\newcommand{\E}{\ensuremath{e}}

\newcommand{\Hz}{\ensuremath{\mathrm{Hz}}}

\newcommand{\omegaMod}{\omega_\mathrm{mod}}

\begin{document}

\title{Atomic magnetic resonance induced by amplitude-, frequency-, or polarization-modulated light}

\author{Z. D. Gruji\'c and A. Weis}

\affiliation{Physics Department, University of Fribourg, CH-1700, Fribourg, Switzerland}

\begin{abstract}
%
In recent years diode laser sources have become widespread and reliable tools in magneto-optical spectroscopy.
In particular, laser-driven atomic magnetometers have found a wide range of practical applications.
More recently, so-called magnetically silent variants of atomic magnetometers have been developed.
While in conventional magnetometers the magnetic resonance transitions between atomic sublevels are phase-coherently driven by a weak oscillating magnetic field, silent magnetometers use schemes in which either the frequency (FM) or the amplitude (AM) of the light beam is modulated.
Here we present a theoretical model that yields algebraic expressions for the parameters of the multiple resonances that occur when either amplitude-, frequency- or polarization-modulated light of circular polarization is used to drive the magnetic resonance transition in a transverse magnetic field.
The relative magnitudes of the resonances that are observed in the transmitted light intensity at harmonic $m$ of the Larmor frequency $\omega_L$ (either by DC or phase sensitive detection at harmonics $q$ of the modulation frequency $\omega_\mathrm{mod}$) of the transmitted light are expressed in terms of the Fourier coefficients of the modulation function.
Our approach is based on an atomic multipole moment representation that is valid for spin-oriented atomic states with arbitrary angular momentum $F$ in the low light power limit.
We find excellent quantitative agreement with an experimental case study using (square-wave) amplitude-modulated (AM) light.

\end{abstract}
\pacs{07.55.Ge, 07.55.Jg, 32.30.Dx, 78.20.Ls, 42.62.Fi}
%
%

\keywords{magnetic resonance; atomic magnetometer; all optical magnetometer; magnetically silent; magneto-optical effects; AM-modulation; FM-modulation; polarization modulation.}

\maketitle

\section*{Introduction}
Magneto-optical spectroscopy of spin-polarized atomic vapors has received a renewed interest in the past decades thanks to the development of solid state diode lasers.
A comprehensive review of of methods and applications of magneto-optical spectroscopy has been given by Budker \emph{et al}~.\cite{BudkerWeis-RevModPhys.74.1153}.
%
%
One of the most prominent applications of spin-polarized atomic vapors prepared by optical pumping with polarized resonance radiation is atomic magnetometry \cite{BudkerRomalisNature}.
Introduced in the late 1950's using discharge lamp pumping, atomic magnetometry has received new interest in the past two decades when diode lasers replaced the lamps \cite{Groeger:LampvsLaser2004}.
Laser pumping has the distinct advantage of allowing multiple sensor arrays to be operated by a single light source \cite{Bison2009:APL19chSystem, PaulNIM} and allows new (magnetically silent) approaches to magnetometry.

Early magnetometers inferred the magnetometry signal of interest directly from the current of a photodetector monitoring the power of the light traversing the atomic medium.
Such magnetometers suffer from low-frequency noise, and the signal/noise ratio, and hence the sensitivity of magnetometers, can be considerably enhanced by using phase-sensitive detection of the photocurrent.
Such lock-in detection requires the application of a suitable modulation to the light-atom interaction.
In the so-called $M_x$-magnetometer scheme \cite{AlexandrovMx,GroegerMx-2007} the modulation is achieved by a weak magnetic field that oscillates at the Larmor frequency and that coherently drives the magnetization associated with the atomic spin polarization around the magnetic field.
The $M_x$-magnetometer is an implementation of optically detected magnetic resonance (ODMR), since the driven spin precession consists, in a quantum picture, of magnetic resonance transitions between magnetic sublevels that are driven by the oscillating field.

In recent years several approaches to so-called magnetically silent (or all-optical) modes of magnetometer operation have been put forward.
These schemes circumvent the application of the oscillating magnetic field, whose implementation may pose technical problems when the magnetometers are operated in harsh environments, such as in ultrahigh vacuum or in the proximity of high voltage \cite{PaulNIM}.
One of the most successful all-optical magnetometry techniques is FM-NMOR (frequency-modulated nonlinear magneto-optical rotation), in which the coherent spin drive (realized by modulation of the laser frequency) is combined with balanced polarimetric detection \cite{Budker:BalRot.PhysRevA.73.053404}.
Amplitude modulation (AM) of the laser intensity is another variant of all optical magnetometry.
It has been implemented in combination with balanced polarimetric detection \cite{GawlikAM} and by using direct power monitoring \cite{JENA:Schultze:12}.
The fact that amplitude modulated resonance light can drive magnetic resonance transitions in the atomic ground state had already been demonstrated by Bell and Bloom in the 1960's, both with circularly \cite{Bell:1961:ODS:PhysRevLett.6.280} and with linearly \cite{BellBloom:Forbidden:PhysRevLett.6.623} polarized light.
Do note, however, that those early experiments did not use phase-sensitive detection.
Yet another, to our knowledge little explored, modulation scheme involves the resonant modulation of the laser polarization.
Only a few examples of polarization modulation have been discussed in the literature \cite{N:Alixandrov:1973, N:Alexandrov-Budker:2005, N:Romalis:2010}.
Below we will refer to polarization modulation as SM (for Stokes modulation, since the acronym PM often refers to phase modulation in the literature).

In this paper we derive algebraic expressions for the magnetic resonance spectra of atomic vapors driven by FM-, AM-, or SM-modulated circularly polarized laser light (modulation frequency $\omega_\mathrm{mod}$).
We analyze the temporal structure of the photodetector signal monitoring the light power after the atomic medium and identify resonant signals modulated at harmonics $q\,\omega_\mathrm{mod}$ of the modulation frequency as well as an unmodulated spectrum of resonances.
When demodulated by a lock-in amplifier tuned to an arbitrary harmonic $q$ of $\omega_\mathrm{mod}$, the magnetic field dependent in-phase and quadrature spectra (for a fixed modulation frequency) show an infinite number of absorptive and dispersive Lorentzian resonances located at multiples $m\,\omega_L$ of the Larmor frequency.
We present algebraic expressions that relate the $q$- and $m$-dependent amplitudes of these resonances to the Fourier coefficients of the modulation function.
Our results are based on an atomic multipole moment approach and are thus applicable to atomic ground states with an arbitrary angular momentum $F$.
The obtained results are valid only in the low light power limit, \emph{i.e.}, in the range where the signal amplitudes grow quadratically with the incident power $P_0$.
In an experimental case study using amplitude modulation (AM) in the low power limit we find an excellent agreement between experimental and theoretical spectra.

Spectra for frequency-modulated (FM) linearly polarized light with polarimetric detection have previously been modeled for a $J=1$ to $J=0$ transition using an algebraic density matrix formalism \cite{Malakyan:PhysRevA.69.013817}.
We are not aware of a related theoretical treatment for AM- or SM-magnetic resonance signals.

\section{Optically induced magnetic resonance}
\label{sec:introduction}
Conventional magnetic resonance is a process in which the orientation of the spin polarization $\vec{S}=\langle\vec{F}\rangle$ of an ensemble of paramagnetic particles (electron, nuclei, atoms)
is changed by a resonant interaction of the associated magnetization
$\vec{M}=\langle\vec{\mu}\rangle\propto\vec{S}$ with a magnetic
field $\vec{b}_1(t)$ oscillating at frequency $\omega_\mathrm{rf}$.
In the case of the atomic ensembles treated here, $\vec{F}$ denotes to the total atomic angular momentum.

\begin{figure}[h]
\begin{center}
\includegraphics[angle=0,width=\textwidth]{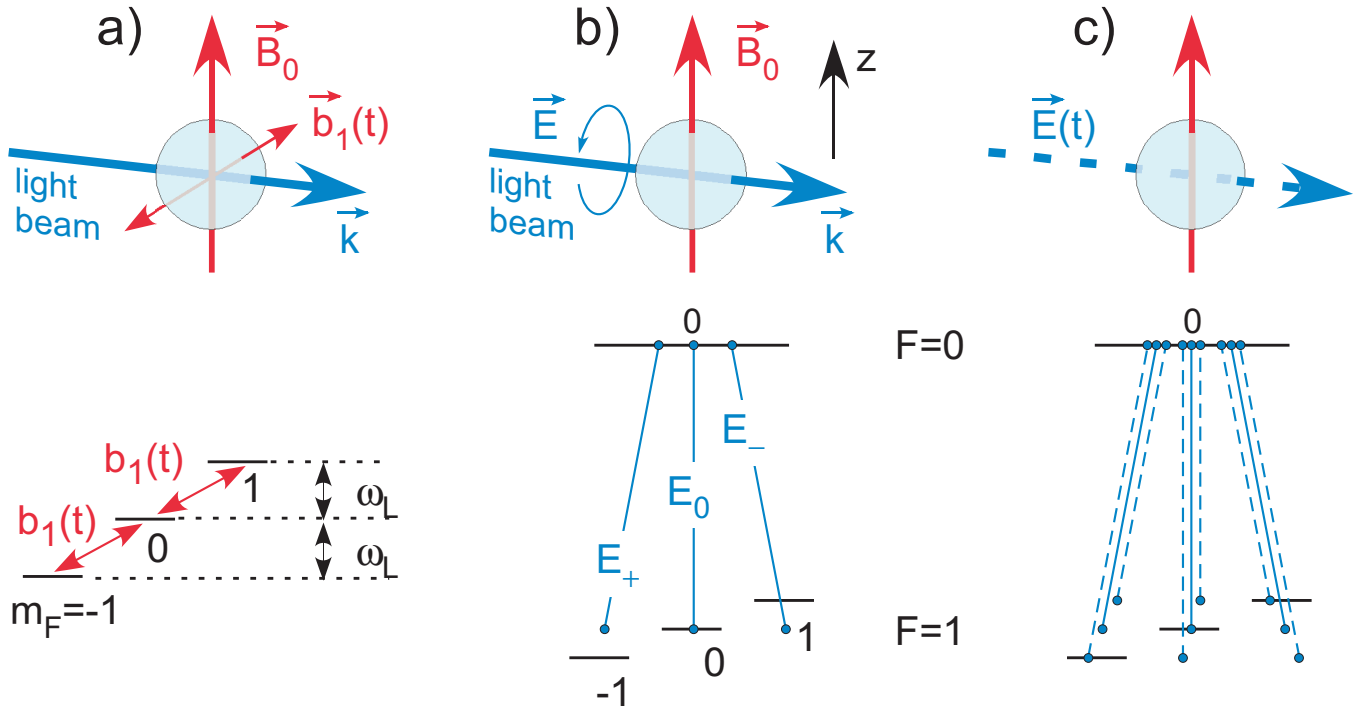}
\caption{(Color online) a) Conventional magnetic resonance: a time varying field
$\vec{b}_1(t)$ induces transitions between magnetic sublevels whose
energies are split by the static field  $\vec{B}_0$.
b) The light beam serves to prepare the spin orientation and to detect the magnetic resonance transition.
$\sigma_\pm$- and $\pi$-polarized components of an unmodulated
circularly polarized light field with quantization axis along
$\vec{B}_0$.
c) Same situation as in (b) when each polarization
component (solid lines) acquires sidebands (dashed lines) due to amplitude modulation that  induce the sublevel transitions.}
\label{fig:sidebands}
\end{center}
\end{figure}

In classical terms, the orientation change of $\vec{S}$ is a Larmor
precession at frequency $\omega_L\propto|\vec{B}_0|$, driven by the
torque $\langle\vec{\mu}\rangle\times\vec{b}_1$.
Magnetic resonance occurs when $\omega_\mathrm{rf}$ matches
$\omega_L$.
In quantum mechanical terms, magnetic resonance is described in terms of magnetic dipole transitions between the magnetic sublevels $\ket{nL_J,F,m_F}$ of the atom
(Fig.~\ref{fig:sidebands}.a), and the transition dynamics are determined by the Hamiltonian
$H=-\vec{\mu}\cdot\vec{b}_1(t)$, with  matrix elements \cite{Nagel:DarkStateMagnetometers1998}
\begin{equation}
\matelem{F, m'_F}{H}{F, m_F}\propto\matelem{nL_J,F,
m'_F}{\vec{\mu}}{nL_J,F, m_F}\cdot\vec{b}_1(t)\,.
\end{equation}
For $\Delta L=0$ transitions, parity conservation requires the
operator driving the transitions to be parity even, \emph{i.e.}, invariant
under space inversion, which is obeyed by $\vec{\mu}$.

In 1961, Bell and Bloom have shown \cite{Bell:1961:ODS:PhysRevLett.6.280} that an intensity-modulated resonant light field with circular polarization induces magnetic
resonance transitions in an atomic ground state when the modulation
frequency $\omegaMod$ matches the ground state's Larmor
frequency $\omega_L$ in a transverse external magnetic field
$\vec{B}_0$.
%
%
The fact that an oscillating electric field can drive $\Delta L=0$, $\Delta F=0$ magnetic resonance transitions seems to be in contradiction with the requirement of parity conservation.
However, the light-induced magnetic resonance transitions can be understood in terms of parity-conserving second-order processes mediated by the
interaction Hamiltonian $H=-\vec{d}\cdot \vec{E}(t)$, as follows.
Consider first an \emph{unmodulated }circularly polarized light
beam, that excites an atomic $F=1\rightarrow F'=0$ transition, in
which the ground state degeneracy is lifted by a transverse magnetic
field $\vec{B}_0$ (Fig.~\ref{fig:sidebands}.b).
With the quantization axis along $\hat{B}_0$, the circularly polarized optical field (oscillating at $\omega$) is
given by
\begin{equation}
\vec{E} = \sum_q E_q \hat{e}_q
=  E_0 \sum_q a_q \hat{e}_q
=E_0 (\frac{1}{2}\hat{e}_++\frac{1}{\sqrt{2}}\,\hat{e}_0+\frac{1}{2}\hat{e}_-)\,,
\end{equation}
where the subscripts $\pm$ and $0$ refer to $\sigma_\pm$- and $\pi$-polarizations, respectively, in a coordinate frame with quantization axis along $\hat{k}$ which drive transitions from all three sublevels.
Because of the energy splitting, only the $0\rightarrow 0$ transition is resonant in the case shown.
When the amplitude of the light is modulated at frequency $\omegaMod$, its Fourier spectrum acquires sidebands that are offset by $\pm n\omegaMod$ from the optical frequency.
In Fig.~\ref{fig:sidebands}.c, we show the carrier $E_0$, oscillating at the optical frequency $\omega$ (solid lines), together with the $n=\pm 1$ sidebands $E_\pm$, oscillating at $\omega \pm \omegaMod$  (dashed lines) for the resonant case where $\omegaMod=\omega_L$.
For simplicity we ignore the higher order sidebands in the present discussion.
These sidebands are responsible for resonances at harmonics of the Larmor frequency (frequency (see also discussion in Sec. 4D of \cite{N:Alexandrov-Budker:2005}).

With this simplification, the carrier $\vec{E}_0$, together with one of the sidebands $\vec{E}_{\pm 1}$ resonantly drive $\Delta m_F=\pm 1$ transitions between adjacent sublevels.
%
%
%
The matrix elements of this second order process can be written in terms of an effective Hamiltonian \cite{NagelWynandsWeis:PhysRevA.58.196} as
\begin{widetext}
\begin{equation}
\matelem{F,m_F'}{H_\mathrm{eff}}{F,m_F}\propto
\sum_{q,q'=0,\pm
1}(-1)^{q+q'}\,\,\matelem{F,m_F'}{d_{-q'}\,d_{-q}}{F,m_F}\,
E_{q'}E_{q}\,. \label{eq:ddEE}
\end{equation}
\end{widetext}
The bilinear form of the dipole operators $d_q$ in \eqref{eq:ddEE} ensures that the matrix elements are parity even and that the effective Hamiltonian $H_\mathrm{eff}$ indeed conserves parity.

Selection rules and relative line strengths for transitions mediated by \eqref{eq:ddEE} were derived in \cite{NagelWynandsWeis:PhysRevA.58.196}.
Moreover, one finds that the two $\sigma_\pm$  polarized
components in Fig.~\ref{fig:sidebands}.c cannot drive $\Delta m_F=2$ transitions
because of destructive quantum interference, thus respecting the
conventional $\Delta m_F=0, \pm 1$ selection rule for magnetic
resonance transitions.
However, when using modulated \emph{linearly polarized} light and a field $\vec{B}_0$, perpendicular to the light polarization, the two sidebands lead to constructive interference, thereby
allowing  $\Delta m_F=2$ transitions to occur.
In this way Bell and Bloom were able to observe the ``forbidden''
$\Delta m_F=2$ magnetic resonance transitions in 1961 \cite{BellBloom:Forbidden:PhysRevLett.6.623} using
amplitude-modulated \emph{linearly }polarized light.

In 1976, Alzetta \emph{et al} \cite{Alzetta:Visualization:1976} devised an elegant method that
allowed the photographic visualization of Bell-Bloom type magnetic
resonance processes induced by polychromatic light fields.
The method has since become known as coherent population trapping (CPT).
%
%
%
We stress an important aspect of CPT spectroscopy:
Since in the ground state the magnetic sublevel coherences excited by bi- or polychromatic light fields may be very long-lived (up to seconds), one has to ensure that the individual Fourier components of the exciting light field have a phase coherence which lives at least as long as the atomic coherence.
With Fourier components produced as sidebands by a modulation technique, the phase coherence is determined by the phase stability of the generator driving the modulator.
However, when the multi-mode light field is produced by a superposition of independent laser sources, special care has to be taken to actively phase-lock the individual optical fields.

\newpage
\section{Magnetic resonance with circularly-polarized modulated light}
\subsection{Experimental geometry}
\label{sec:ExpGeometry}
\begin{figure}[b]
\begin{center}
\includegraphics[angle=0,width=\columnwidth]{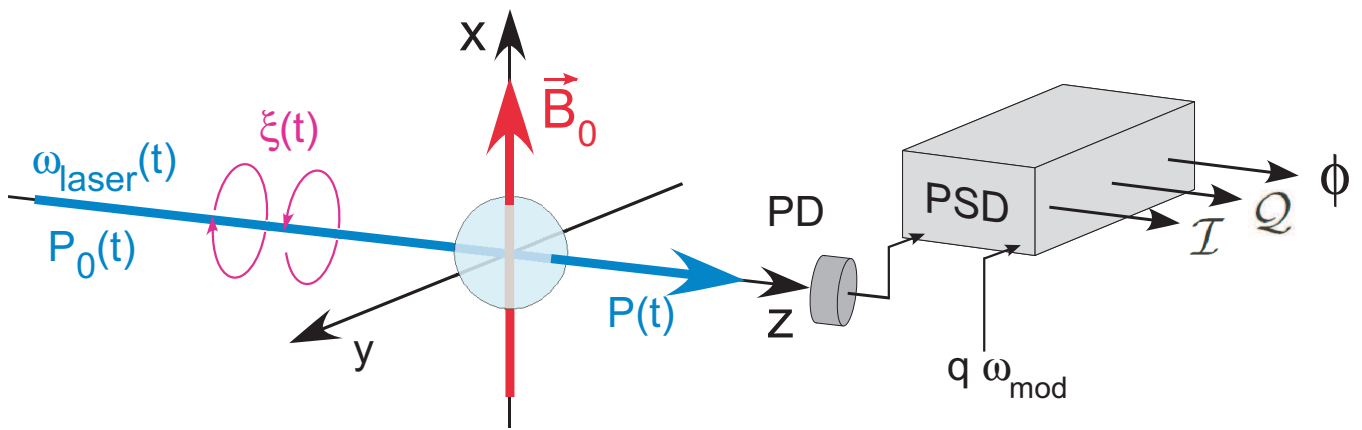}
\caption{(Color online) Experiments addressed in this paper. A circularly polarized
resonant light beam traverses an atomic medium exposed to a static
magnetic field $\vec{B}_0$.
Either the power $P_0$, the frequency $\omega_\mathrm{laser}$, or the polarization (helicity) $\xi$ of the light is
modulated at frequency $\omegaMod$ and a phase sensitive
detector (PSD), tuned to $q\omegaMod$, extracts the in-phase component ($\mathcal{I}$), the quadrature component
($\mathcal{Q}$),  and the phase ($\phi$) of the signal from the photodiode (PD) detecting the transmitted modulated power $P(t)$.} \label{fig:expsetup}
\end{center}
\end{figure}

Figure \ref{fig:expsetup} shows the geometry of the experiments
discussed in this paper.
A circularly polarized laser beam, resonant with an $F\rightarrow
F'$ transition, traverses an atomic vapor cell of length $L$ that is
exposed to a transverse static magnetic field $\vec{B}_0$.
The power $P(t)$ of the light transmitted  through the cell is measured
by a photodetector.
For suppression of technical noise, one may wish to use a balanced polarimeter detecting
alterations of the light polarization rather than merely detecting
the light power.
Such extensions of the method will not be addressed here.

We will discuss three distinct experiments, in which a given
property, \emph{viz.}, the power $P_0(t)$, the frequency detuning
$\delta\omega(t)$ from the atomic transition, or the
helicity $\xi(t)$ of the incident light field is subject to a
periodic modulation at frequency $\omegaMod$.
The modulated property will imprint a characteristic periodic
modulation at $\omegaMod$, or harmonics $q$ thereof, onto the
power of the transmitted beam, with amplitude(s) and phase shift(s)
that depend on the detuning of the Larmor frequency
$\omega_L=\gamma_F\,|\vec{B}_0|$ from $q\,\omegaMod$,
where $\gamma_F$ is the gyromagnetic ratio of the polarized ground
state $F$.

In the experiments, the time dependent photodetector signal $P(t)$ is
analyzed by a phase-sensitive detector (PSD), referenced to
$\omegaMod$ or its $q$-th harmonic.
At each demodulation frequency q\,$\omegaMod$, one
observes a  series of resonances at multiples $m\,\omega_L$ ($m$, arbitrary integer) of the
Larmor frequency.
The aim of the present paper is the derivation of algebraic
expressions for the amplitudes $a_{q,m}$ and $d_{q,m}$ of the in-phase
and quadrature components of $P(t)$.

\subsection{Light transmission through a spin-polarized vapor}
\label{sec:Lambert-Beer}
The light power $P$ transmitted by an unpolarized optically thin atomic vapor of length $L$ is given by
\begin{equation}
P=P_0\,\E^{-\kappa(\delta\omega)\,L}\approx P_0-P_0\,\kappa(\delta\omega)\,L\,,
\end{equation}
where
\begin{equation}
\kappa(\delta\omega)=\kappa_0\,D(\delta\omega)
\end{equation}
is the optical absorption coefficient, parametrized in terms of the
peak absorption coefficient $\kappa_0$ and a spectral lineshape
function $D(\delta\omega)$, typically a Doppler or Voigt profile with $D(0)=1$, that depends on the detuning $\delta\omega=\omega_\mathrm{laser}-\omega_0$ of the laser frequency $\omega_\mathrm{laser}$ from the atomic resonance frequency $\omega_0$.

When the medium is spin polarized, the peak absorption coefficient for light with circular polarization $\sigma_\xi$ has to be replaced by
\begin{equation}
\kappa_0\rightarrow\kappa_0\left[1-\alpha_{F,F'}\,\xi\,S_z-\beta_{F,F'}\,A_{zz}\right]\,,
\end{equation}
where
\begin{equation}
S_z=\frac{1}{F}\,\sum_{m_F=-F}^{F}m_F\,p_{m_F} \label{eq:SzDef}
\end{equation}
and
\begin{equation}
A_{zz}=\frac{1}{F(2F-1)}\,\sum_{m_F=-F}^{F}\left[3m_F^2-F(F+1)\right]\,p_{m_F}
\label{eq:AzzDef}
\end{equation}
are the vector polarization (orientation) and tensor polarization
(alignment) of the medium, respectively, with $p_{m_F}$ being the normalised
sublevel populations $\sum p_{m_F}=1$.
Both $S_z$ and $A_{zz}$ are defined here to be normalized to unity when the system is in the stretched state defined by $p_{m_F}=\delta_{m_F,F}$.
The coefficients $\alpha_{F,F'}$ and $\beta_{F,F'}$ depend on
the angular momenta $F, F'$ of the states coupled by the optical
transition.

Optical pumping with circularly polarized light produces both
orientation and alignment in the ground state.
In order not to overcharge the present paper we will consider only
orientation contributions by setting $\beta_{F,F'}=0$.
As discussed at the end of the paper, we have in fact observed weak signal components that can be assigned to alignment contributions.
Since these components are spectrally resolved from the orientation contributions, they will not be addressed here.
With the above restrictions the transmitted power is given by
\begin{equation}
P=\left[1-\kappa_0L\,D(\delta\omega)\right]\,P_0+\alpha\kappa_0L\,D(\delta\omega)\,\xi\,S_z\,P_0\,.
\label{eq:PoutBasic}
\end{equation}
We will not address the dependence of $\alpha_{F,F'}$ on $F$ and $F'$, and drop the indices in consequence.
The combinations $\kappa_0 L$ and $\alpha \kappa_0 L$ can be seen as experimental parameters which can be determined empirically.

\subsection{Modulation schemes}
\label{sec:ModSchemes}
We address the following three modulation schemes:
\begin{itemize}
%
\item \textbf{Amplitude modulation (AM):}
The laser frequency is set to resonance,
$D(\delta\omega=0)=1$, the polarization is fixed to
$\xi=+1$, and the incident light power is modulated by an arbitrary
periodic time dependent function according to
$P_0(t)~=~P_0\,f_{\omegaMod}^\mathrm{AM}(t)$.
The corresponding time dependence of the detected power then reads
\begin{widetext}
\begin{align}
P^\mathrm{AM}(t)&=\left(1-\kappa_0L\right)\,P_0(t)+\alpha\kappa_0L\,S_z^\mathrm{AM}(t)\,P_0(t)\\
&=\left(1-\kappa_0L\right)\,P_0\,f_{\omegaMod}^\mathrm{AM}(t)+\alpha\kappa_0L\,P_0\,S_z^\mathrm{AM}(t)\,f_{\omegaMod}^\mathrm{AM}(t)\\
&\equiv
A^\mathrm{AM}+B^\mathrm{AM}\,f_{\omegaMod}^\mathrm{AM}(t)+C^\mathrm{AM}\,S_z^\mathrm{AM}(t)\,f_{\omegaMod}^\mathrm{AM}(t)\,.
\label{eq:AMPoft}
\end{align}
\end{widetext}
%

\item \textbf{Frequency modulation (FM):}
The incident power is fixed to $P_0$, the helicity of the light
polarization is fixed to $\xi=+1$, and the laser detuning
$\delta\omega(t)$ is periodically modulated.
%
The corresponding time dependence of the detected power reads
\begin{widetext}
\begin{align}
P^\mathrm{FM}(t)&=\left[1-(\kappa_0L)\,D\left(\delta\omega(t)\right)\right]\,P_0
+(\alpha\,\kappa_0L)\,P_0\,S_z^\mathrm{FM}(t)\,D\left(\delta\omega(t)\right)\,.
\label{eq:AMPoft3}
\end{align}
\end{widetext}
The modulation function can be modeled by replacing $D\left(\delta\omega(t)\right)$ by a periodic function $f_{\omegaMod}^\mathrm{FM}(t)$, with $0\le
f_{\omegaMod}^\mathrm{FM}\le 1$.
With this choice, the transmitted power can be written as
\begin{widetext}
\begin{align}
P^\mathrm{FM}(t)&=P_0-(\kappa_0L\,P_0)\,f_{\omegaMod}^\mathrm{FM}(t)+(\alpha\,\kappa_0L\,P_0)\,S_z^\mathrm{FM}(t)\,f_{\omegaMod}^\mathrm{FM}(t)\\
&\equiv
A^\mathrm{FM}+B^\mathrm{FM}\,f_{\omegaMod}^\mathrm{FM}(t)+C^\mathrm{FM}\,S_z^\mathrm{FM}(t)\,f_{\omegaMod}^\mathrm{FM}(t)\,.
\label{eq:AMPoft2}
\end{align}
\end{widetext}

\item \textbf{Polarization modulation (SM):}
The laser frequency is set to resonance,
$D(\delta\omega=0)=1$, the incident power is fixed to
$P_0$, and the helicity (degree of circular polarization) of the
light is periodically modulated as
$\xi(t)=f_{\omegaMod}^\mathrm{SM}(t)$, with
$|f_{\omegaMod}^\mathrm{SM}|\le 1$.
The corresponding time dependence of the detected power reads
\begin{widetext}
\begin{align}
P^\mathrm{SM}(t)&=(1-\kappa_0L)\,P_0+(\alpha\,\kappa_0L\,P_0)\,S_z^\mathrm{SM}(t)\,f_{\omegaMod}^\mathrm{SM}(t)\\
&\equiv
A^\mathrm{SM}+B^\mathrm{SM}\,f_{\omegaMod}^\mathrm{SM}(t)+C^\mathrm{SM}\,S_z^\mathrm{SM}(t)\,f_{\omegaMod}^\mathrm{SM}(t)\,.
\label{eq:PMPoft3}
\end{align}
\end{widetext}
\end{itemize}
We see that all three types of experiments (TOE) can be parametrized in terms of distinct time independent and time dependent terms of the general form
\begin{equation}
P^\mathrm{TOE}(t)=A^\mathrm{TOE}+B^\mathrm{TOE}\,f_{\omegaMod}^\mathrm{TOE}(t)+C^\mathrm{TOE}\,S_z^\mathrm{TOE}(t)\,f_{\omegaMod}^\mathrm{TOE}(t)\,.
\label{eq:PTOE}
\end{equation}
The parameters $A$, $B$, $C$ for amplitude-, frequency-, and polarization-modulation are summarized in Table~\ref{tab:ABC}.

%
\begin{center}
\begin{table}[h]
\begin{tabular}{cccccc}
TOE&\;&$A^\mathrm{TOE}$ & $B^\mathrm{TOE}$ & $C^\mathrm{TOE}$ & $f_{\omegaMod}^\mathrm{TOE}$(t)\\
\hline
AM&\;&0&  $\left(1-\kappa_0L\right)\,P_0$ & $~~\alpha\,\kappa_0L\,P_0~~$ & $f_{\omegaMod}^\mathrm{AM}\in [0,1]$ \\
FM&\;& $P_0$& $-\kappa_0L\,P_0$ & $\alpha\,\kappa_0L\,P_0$ & $f_{\omegaMod}^\mathrm{FM}\in [0,1]$ \\
SM&\;& $\left(1-\kappa_0L\right)\,P_0$& 0 & $~~\alpha\,\kappa_0L\,P_0~~$ & $f_{\omegaMod}^\mathrm{SM}\in [-1,1]$
\end{tabular}
\caption{Characteristic parameters $A^\mathrm{TOE}$,
$B^\mathrm{TOE}$, and $C^\mathrm{TOE}$ for experiments with
amplitude- (AM), frequency- (FM), and polarization- (SM) modulated
light, respectively.
The last column gives the lower and upper bounds of the modulation
functions $f_{\omegaMod}^\mathrm{TOE}$(t) for achieving a maximal contrast of the system's response.\label{tab:ABC}}
\end{table}
\end{center}
%
We note the following facts:
\begin{itemize}
\item
The time-independent term $A^\mathrm{TOE}$ gives no contribution to the lock-in signals.
\item
The term $B^\mathrm{TOE}$ has the same Fourier spectrum as the time dependent modulation, but contains no magnetic field dependent quantities.
In AM and FM experiments it will lead to a field independent
background that is, for an optically thin medium, $\kappa_0L\ll1$,
substantially larger in AM experiments than in FM experiments, while
in SM experiments it is absent.
\item
The $C^\mathrm{TOE}$ term leads to a richer spectrum because of the
mixing of frequencies of its two time dependent contributions
$S_z(t)$ and $f_{\omegaMod}^\mathrm{TOE}(t)$.
We  note that only the $C^\mathrm{TOE}$ term depends---via $S_z(t)$--- on the magnetic field, while the  $A^\mathrm{TOE}$  and  $B^\mathrm{TOE}$  terms  form a signal background that influences the contrast and the signal/noise ratio of the magnetic resonance structures.
\end{itemize}
In Section \ref{sec:Szoft} we will first derive algebraic expressions
that relate the time dependent  spin orientation $S_z^\mathrm{TOE}(t)$ to the specific drive function $f_{\omegaMod}^\mathrm{TOE}(t)$, and in section \ref{sec:lock-in} we
will then derive and discuss the complete Fourier spectra of the signals
$P^\mathrm{TOE}(t)$.

\section{Spin orientation $S_z(t)$ under periodic modulation}
\label{sec:Szoft}
As stated above, we will not address alignment contributions to the atomic polarization and describe the latter only in terms of its vector polarization (orientation) $\vec{S}$.
The dynamics of $S_z(t)$, \emph{i.e.}, the only polarization component that contributes to the signals, is governed by the Bloch equations
\begin{equation}\label{eq:BlochVec}
    \dot{\vec{S}}=\vec{S}\times\vec{\omega}_L-\gamma\,\vec{S}+\Gamma_p^\mathrm{TOE}(t)\,\hat{k}\,,
\end{equation}
whose components read
\begin{align}
\dot{S}_x&=-\gamma\,S_x\label{eq:uEqMot}\\
\dot{S}_y&=+\omega_L\,S_z-\gamma\,S_y\label{eq:vEqMot}\\
\dot{S}_z&=-\omega_L\,S_y-\gamma\,S_z+\Gamma_p^\mathrm{TOE}(t)\label{eq:wEqMot}\,,
\end{align}
where we have assumed that the longitudinal and transverse
relaxation rates are identical $\gamma_1=\gamma_2\equiv\gamma$, and
where $\Gamma_p^\mathrm{TOE}(t)$ is a source term that describes the rate at
which longitudinal orientation $S_z$ is produced by optical pumping.
We note that the pumping rate $\Gamma_p(t)$ is proportional to the
product of the incident power $P_0$, the light helicity $\xi$, and
the optical lineshape $D(\delta\omega)$.
The modulation of any of these quantities thus yields a modulation
of the pumping (and probing) rate, that can be parametrized as
\begin{equation}
\Gamma_p^\mathrm{TOE}(t)=\gamma_p\,f_{\omegaMod}^\mathrm{TOE}(t)\,,
\end{equation}
where $f_{\omegaMod}^\mathrm{TOE}(t)$ is the modulation
function which varies periodically within the bounds listed in Table~\ref{tab:ABC}.
The pumping rate amplitude can be related to the light power $P_0$
(or the light intensity $I_A$) by introducing a saturation parameter $G$, defined as
\begin{equation}
G \equiv \frac{\gamma_p}{\gamma}\equiv \frac{P_0}{P_s}\equiv\frac{I}{I_s}\,,
\end{equation}
where $P_s$ and $I_s$ are the saturation power and saturation intensity, respectively.

We note that the Bloch equations above, and hence the
solutions below, are only valid in the low power
approximation $P_0\ll P_S$, \emph{i.e.}, $\gamma_p\ll\gamma$, which
expresses the fact that less than one optical pumping
(absorption/fluorescence) cycle occurs during the lifetime
$\gamma^{-1}$ of the ground state polarization.

\subsection{Monochromatic modulation}
Using Wolfram Mathematica 8.0 \cite{Mathematica}, one can show that the Bloch equation for $S_z(t)$ driven by a
monochromatic modulation around a DC offset value
\begin{equation}
\Gamma_p^{\mathrm{AM,FM}}(t)=\frac{\gamma_p}{2}\left[1+\cos\left(\omegaMod t\right)\right]\,.
\label{eq:AMFMmod}
\end{equation}
has a time dependent solution of the form
\begin{equation}
S_z^{\mathrm{AM,FM}}(t)=\frac{\gamma_p}{2}\left[T(t)+R(t)\right] \,,
\end{equation}
with
\begin{widetext}
\begin{align}
T^{\mathrm{AM,FM}}(t)=&\left[\frac{2 \,\omega_L}{\omega_L^2+\gamma^2}  +  \frac{\omega_L-\omegaMod}{(\omega_L-\omegaMod)^2+\gamma^2}  + \frac{\omega_L+\omegaMod}{(\omega_L+\omegaMod)^2+\gamma^2}\right]\,\sin(\omega_L t)\,\E^{-\gamma t}\nonumber\\
-&\left[\frac{2\,\gamma}{\omega_L^2+\gamma^2}  +  \frac{\gamma}{(\omega_L-\omegaMod)^2+\gamma^2}  +  \frac{\gamma}{(\omega_L+\omegaMod)^2+\gamma^2}\right]\,\cos(\omega_L t)\,\E^{-\gamma t}\,,\\
R^{\mathrm{AM,FM}}(t)=&\frac{2\,\gamma}{\omega_L^2+\gamma^2} \nonumber\\
+&\left[ \frac{\gamma}{(\omegaMod-\omega_L)^2+\gamma^2}  +  \frac{\gamma}{(\omegaMod+\omega_L)^2+\gamma^2}\right]\,\cos(\omegaMod t)\nonumber\\
+&\left[\frac{\omegaMod-\omega_L}{(\omegaMod-\omega_L)^2+\gamma^2}
+\frac{\omegaMod+\omega_L}{(\omegaMod+\omega_L)^2+\gamma^2}\right]\,\sin(\omegaMod
t)
\,. \label{eq:SSsolutions}
\end{align}
\end{widetext}

The function $T^{\mathrm{AM,FM}}(t)$ is a damped transient, so that for $t\gg\gamma^{-1}$ $S_z(t)$ shows a steady-state oscillation given by
\begin{widetext}
\begin{align}
S_z^{\mathrm{AM,FM}}(t)&=\frac{\gamma_p}{2}\,R(t)^{\mathrm{AM,FM}}\\
&=\frac{G}{2} \mathcal{H}(\omega_L)
+\frac{G}{2} \left[\mathcal{A}(\omega_L)+\mathcal{A}(-\omega_L)\right]\,\cos(\omegaMod t)\\
&\phantom{=\frac{G}{2} \mathcal{H}_0(\omega_L)}+\frac{G}{2} \left[\mathcal{D}(\omega_L)+\mathcal{D}(-\omega_L)\right]\,\sin(\omegaMod
t)\, , \label{eq:SzAMFM}
\end{align}
\end{widetext}
with resonance line shapes
%
\begin{align}
\mathcal{H}(\omega_L)&=\frac{2\,\gamma^2}{\omega_L^2+\gamma^2}\\
\mathcal{A}(\omega_L)&=\frac{\gamma^2}{(\omegaMod-\omega_L)^2+\gamma^2}\\
%
\mathcal{D}(\omega_L)&=\frac{\gamma\,(\omegaMod-\omega_L)}{(\omegaMod-\omega_L)^2+\gamma^2}\,.
\end{align}
%
%
The spin polarization thus contains an unmodulated DC Lorentzian (Hanle) resonance centred at $\omega_L=0$, as well as absorptive and dispersive Lorentzians, centred at $\omega_L=\pm\omegaMod$.

We note that the DC term in the pumping rate \eqref{eq:AMFMmod} occurs only in the AM and FM schemes, while the SM modulation function (normalized to the same peak-peak modulation amplitude)
\begin{equation}
\Gamma_p^{\mathrm{SM}}(t)=\frac{\gamma_p}{2}\,\cos\left(\omegaMod t\right)
\label{eq:PMmod}
\end{equation}
has no DC part.

The steady-state Bloch oscillations in that case are given by
\begin{widetext}
\begin{equation}
S_z^{\mathrm{SM}}(t)
=\frac{G}{2}\,\{\left[\mathcal{A}(\omega_L)+\mathcal{A}(-\omega_L)\right]\,\cos(\omegaMod
t)
+\left[\mathcal{D}(\omega_L)+\mathcal{D}(-\omega_L)\right]\,\sin(\omegaMod
t)\}\,, \label{eq:SzPM}
\end{equation}
\end{widetext}
and show no Hanle resonance in the unmodulated DC signal.
\subsection{Arbitrary periodic modulation}
We consider next an arbitrary symmetric ($g_{m}=g_{-m}$) periodic modulation that
can be represented in terms of its $cosine$-Fourier series
\begin{equation}
\Gamma_p^\mathrm{TOE}(t)
=\gamma_p\,f_{\omegaMod}^\mathrm{TOE}(t)
=\gamma_p\,\sum_{m=-\infty}^\infty
g_{m}\,\cos\left(m\,\omegaMod t\right)\,.
\label{eq:ArbMod}
\end{equation}
Since the Bloch equations are linear in $\gamma_p$, they can be solved for each Fourier component  $\cos\left(m\,\omegaMod t\right)$ independently, yielding
\begin{widetext}
\begin{align}
S_z^{(m)}(t)
&=g_m\,G\,\left\{\left[\mathcal{A}_{m}(\omega_L)+\mathcal{A}_{m}(-\omega_L)\right]\cos(m \omegaMod
t)
+\left[\mathcal{D}_{m}(\omega_L)+\mathcal{D}_{m}(-\omega_L)\right]\sin(m\omegaMod
t)\right\}\nonumber\\
&=g_m\,G\,\left\{\left[\mathcal{A}_{m}(\omega_L)+\mathcal{A}_{-m}(\omega_L)\right]\cos(m \omegaMod
t)
+\left[\mathcal{D}_{m}(\omega_L)-\mathcal{D}_{-m}(\omega_L)\right]\sin(m\omegaMod
t)\right\}\,, \label{eq:SzAMFM2}
\end{align}
\end{widetext}
with
%
\begin{align}
\mathcal{A}_{m}(\omega_L)&=\frac{\gamma^2}{(m\omegaMod-\omega_L)^2+\gamma^2}\,,\\
%
\mathcal{D}_{m}(\omega_L)&=\frac{\gamma(m\omegaMod-\omega_L)}{(m\omegaMod-\omega_L)^2+\gamma^2}\,.
\end{align}
%
Note that we have replaced the experiment indicating superscript $TOE$ on $S_z(t)$ by the order $m$ of the resonance behaviour of the Fourier coefficient, and have added the subscript $m$ to $\mathcal{A}(\omega_L)$ and $\mathcal{D}(\omega_L)$ to denote the Fourier component at $m \omegaMod$.
By a proper choice of the coefficients $g_m$, the expressions can be applied to all three types of experiments.

Summing all Fourier components, we find that the time dependent spin polarization is given by
\begin{widetext}
\begin{equation}
S_z(t)=\sum_{m=-\infty}^\infty S_z^{(m)}(t)=2\,G\sum_{m=-\infty}^\infty\,g_m\,\left[\mathcal{A}_{m}(\omega_L)\,\cos(m
\omegaMod t)
+\mathcal{D}_{m}(\omega_L)\,\sin(m\omegaMod t)\right]\,. \label{eq:SzArbMod}
\end{equation}
\end{widetext}
The factor $2$ in the last expression originates from the symmetries $\mathcal{A}_{-m}=\mathcal{A}_{m}$ and $\mathcal{D}_{-m}=-\mathcal{D}_{m}$ of the lineshape functions.
The Hanle resonance of \eqref{eq:SzAMFM} is now explicitly contained as the $m=0$ term in the sum since $\mathcal{H}=\mathcal{A}_0+\mathcal{A}_{-0}$.
%
\section{The lock-in signals} 
\label{sec:lock-in}
In Appendix A we show that the transmitted power (photodiode signal) contains time independent and time dependent contributions
\begin{equation}
P^\mathrm{TOE}(t)=\mathcal{B}_\mathrm{DC}^\mathrm{TOE}
+\sqrt{2}\sum_{q=1}^\infty\,\mathcal{I}^\mathrm{TOE}_\mathrm{q}\,\cos\left(q\,\omegaMod t\right)
+\sqrt{2}\sum_{q=1}^\infty\,\mathcal{Q}^\mathrm{TOE}_\mathrm{q}\,\sin\left(q\,\omegaMod t\right)\,.
\label{eq:2}
\end{equation}
The time-independent (DC) signal is given by
\begin{widetext}
\begin{equation}
\mathcal{B}_\mathrm{DC}^\mathrm{TOE}=A^\mathrm{TOE}+g_0\,B^\mathrm{TOE}
+G\,C^\mathrm{TOE}\sum_{m=-\infty}^{\infty}g_m^2\,\mathcal{A}_m(\omega_L)\,,
\end{equation}
\end{widetext}
which represents an infinite series of absorptive Lorentzians, centered at $\omega_L=m\,\omegaMod$, respectively, that are superposed on a field independent background $A^\mathrm{TOE}+g_0\,B^\mathrm{TOE}$.
The constants $A^\mathrm{TOE}$, $B^\mathrm{TOE}$, and $C^\mathrm{TOE}$ are given in Table \ref{tab:ABC} for the different types of experiments.
We note that for $TOE$=$AM$, the $m=0$ term represents the magnetic resonance described in the early work of Bell and Bloom \cite{Bell:1961:ODS:PhysRevLett.6.280}.

The photodiode signal further contains (periodic) time-dependent components that oscillate in-phase  and in quadrature  with the fundamental and higher harmonics of the modulation frequency $\omegaMod$.
$\mathcal{I}^\mathrm{TOE}_\mathrm{q}(\omega_L)$ and
$\mathcal{Q}^\mathrm{TOE}_\mathrm{q}(\omega_L)$ represent the rms amplitudes of these signals, when extracted by a lock-in amplifier (Fig.~\ref{fig:expsetup}) referenced by $\cos{q\omegaMod}$, respectively:
\begin{widetext}
\begin{equation}
\mathcal{I}^\mathrm{TOE}_\mathrm{q}(\omega_L)=h_q\,
+\sum_{m=-\infty}^\infty a_{q,m}\,\mathcal{A}_{m}(\omega_L)\,,
\label{eq:IPq0}
\end{equation}
\end{widetext}
with
\begin{widetext}
\begin{equation}
h_q=\sqrt{2}\,g_q\,B^\mathrm{TOE}\quad\text{and}\quad a_{q,m}=\sqrt{2}\,G\,C^\mathrm{TOE}\,\,g_m\,(g_{q-m}+g_{q+m})\,.
\label{eq:IPq02}
\end{equation}
\end{widetext}
The corresponding quadrature signals read
\begin{widetext}
\begin{equation}
\mathcal{Q}^\mathrm{TOE}_\mathrm{q}=\sum_{m=-\infty}^\infty
\,d_{q,m}\mathcal{D}_{m}(\omega_L),
\quad\text{with}\quad
d_{q,m}=\sqrt{2}\,G\,C^\mathrm{TOE}\,\,g_m\,(g_{q-m}-g_{q+m})\,.
\label{eq:QUq0}
\end{equation}
\end{widetext}
At each demodulation harmonic $q$, one thus observes an infinite series of absorptive and dispersive Lorentzians, centered at $\omega_L=\pm m\,\omegaMod$.
The absorptive resonances of the in-phase spectrum have amplitudes given by $a_{q,m}$ which are expressed in terms of a type-of-experiment specific constant $C^\mathrm{TOE}$, the peak optical pumping rate $\gamma_p$ (itself proportional to the incident laser power $P_0$), and a simple algebraic function of the Fourier coefficients $g_i$ of the specific modulation function $f^\mathrm{TOE}(t)$ of similar composition.
The quadrature spectrum consists of dispersive Lorentzians of amplitudes $d_{q,m}$.

We note that the in-phase resonance spectrum is superposed on a magnetic field independent background of amplitude $h_q$, which vanishes for polarization modulation ($TOE=SM$), and is reduced by a factor $\alpha\kappa_0L$ in $FM$ experiments compared to $AM$ experiments.
The quadrature spectrum is background-free at all demodulation harmonics and for all three types of experiments.
We further note that the in-phase spectrum contains absorptive zero-field ($m$=0) Hanle resonances at all demodulation harmonics $q\,\omegaMod$, which have no dispersive counterparts in the quadrature signals since $d_{q,0}$=0.

The linear zero-crossings at the centers of the dispersive resonances offer a convenient discriminator signal for magnetometers in which active feedback is used to stabilize the Larmor frequency $\omega_L$ to the modulation frequency $\omegaMod$ (or vice-versa).
%


\section{Magnetic resonance induced by amplitude modulated light}
%
\subsection{Experiments}
In order to illustrate how well equations
\eqref{eq:IPq0}--\eqref{eq:QUq0} describe experimental spectra, we
present the result of a case study using amplitude modulated light.
Related experiments were reported in the literature \cite{GawlikAM, JENA:Schultze:12}.
Our experiments were done in a paraffin-coated Cs vapor cell,
using a laser beam whose frequency was actively stabilized to the
$4\rightarrow 3$ hyperfine transition of the $D_1$-line.
Following the procedure outlined in \cite{Castagna:PhysRevA.84.053421}, we choose the light
power to be sufficiently low ($1.4~\mu W$) to ensure the validity of
the theoretical model.
The experiments were carried out in a threefold $\mu$-metal shield,
in which the power of the circularly polarized laser beam traversing
the cell was recorded by a photodiode.
We used a compact fiber-coupled sensor---similar to the one described in \cite{castagna:Cells:2009}---that contained the polarization optics, the vapor cell and the photodiode, and that was mounted inside of a long solenoid producing the transverse magnetic field.


The laser intensity was given a square-wave on/off modulation with a
50/50 duty cycle using an acousto-optic modulator.
%
The photocurrent was amplified by a transimpedance amplifier and
analyzed by a Zurich Instruments (model HF2LI) lock-in amplifier,
which allowed the simultaneous extraction of the in-phase and
quadrature components at six selected harmonics of its reference
frequency.

%
\subsection{Analysis and discussion}
\begin{figure}[b]
\begin{center}
\includegraphics[angle=0,width=\columnwidth]{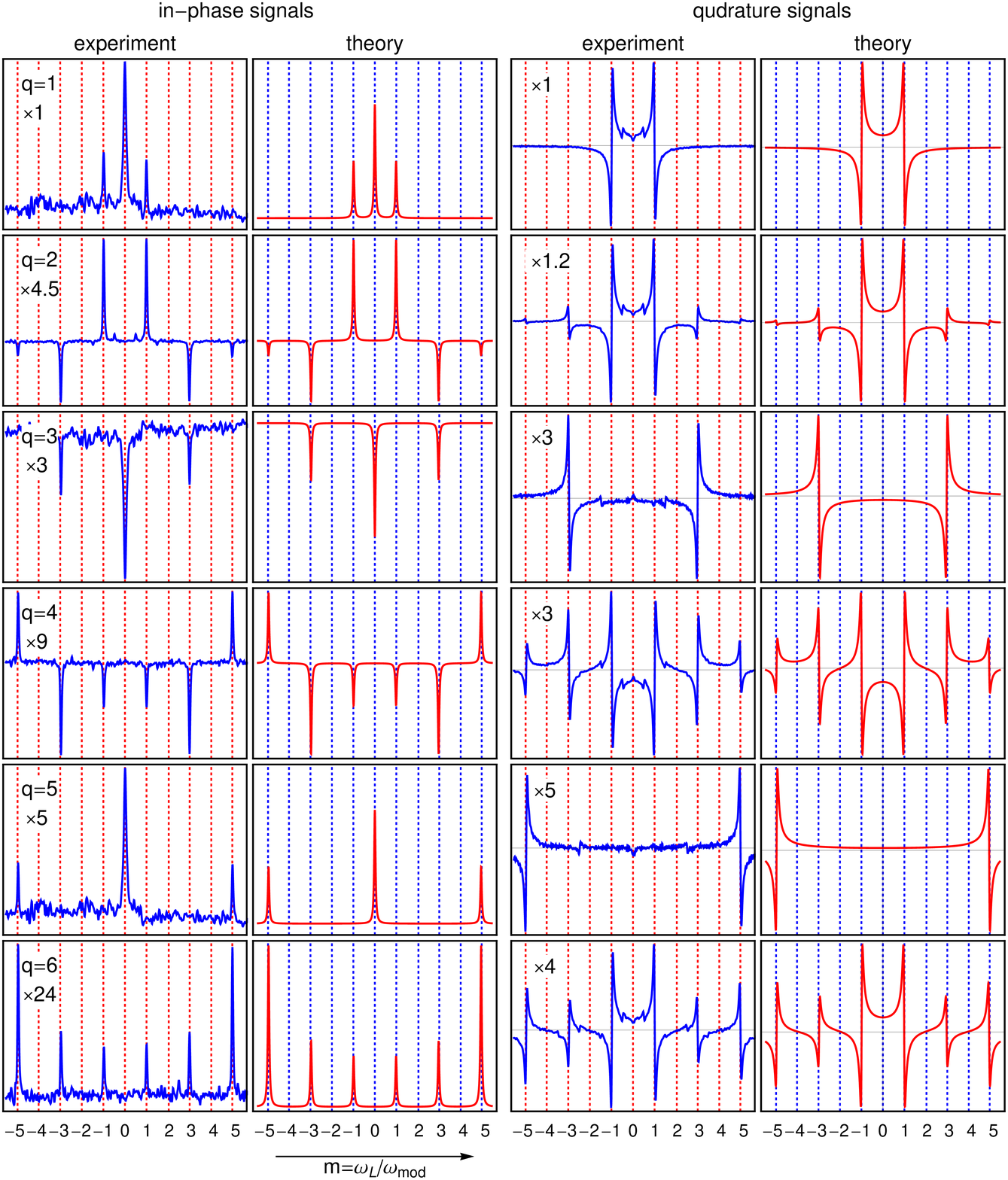}
\caption{Comparison of experimental and theoretical magnetic resonance spectra excited by amplitude-modulated circularly polarized light (50\% duty cycle) in a transverse magnetic field (Larmor frequency $\omega_L$ ) .
The  modulation frequency  $\omega_{mod}=(2\pi)127~\Hz$ was kept constant while the magnetic field was scanned.
The magnetic field amplitude is represented in units of the modulation frequency ($m=\omega_L/\omegaMod$).
The left and right parts of the figure show, respectively, the lock-in extracted in-phase $\mathcal{I}$ and quadrature $\mathcal{Q}$ components of the photocurrent monitoring the transmitted modulated light intensity.
The parameter $q$ denotes the harmonic of $\omegaMod$ at which the signal was demodulated.
In each column, all spectra are normalized to the amplitude of the $q=1, m=1$ resonance in that column.}
\label{fig:AMresonances}
\end{center}
\end{figure}

The results are shown in Fig.~\ref{fig:AMresonances}, together with the theoretical
prediction based on \eqref{eq:IPq0}--\eqref{eq:QUq0}, evaluated with the Fourier
coefficients
\begin{equation}\label{eq:FTSquareWave}
    g_0=\frac{1}{2}\quad\text{and}\quad g_{k\neq 0}=\frac{1}{\pi}\,\frac{\sin(k\,\frac{\pi}{2})}{k}
\end{equation}
of the symmetric, $f_\mathrm{mod}(t)=f_\mathrm{mod}(-t)$, square wave modulation function.
The only post-treatment applied to the recorded data was the subtraction of
the DC offset of the in-phase components and the scaling of all
experimental data $\mathcal{Q}_q$, and $\mathcal{I}_q$ (after the mentioned offset
subtraction) by a \emph{one common} multiplicative factor, chosen such that the theoretical amplitudes of the dispersive resonance in the first harmonic ($q=1$) spectrum match the amplitudes of the corresponding experimental spectrum.
One sees that this single scaling factor yields an excellent agreement between the experimental spectra and the theoretical predictions for the whole range of investigated $q$ and $m$ values.

In Table~\ref{tab:comparison} we  present a quantitative comparison of the predicted and measured peak-peak amplitudes of the dispersive resonances, since the quadrature resonances have a superior signal/noise ratio (SNR).
The experimental zero in the Table~\ref{tab:comparison} means that under our experimental conditions a resonance was not observed.
The poorer SNR of the in-phase signals arises from the background signal $B^\mathrm{TOE}$ that is proportional to the incident laser power $P_0$, so that power fluctuations of $P_0$ produce noise on the in-phase components that surpasses the noise on the quadrature components by a factor on the order of $B^\mathrm{AM}/C^\mathrm{AM}\approx (\alpha\kappa_0L)^{-1}$.
We note that this excess noise factor reduces to $\alpha^{-1}$ in the case of $FM$ experiments, and that it is unity for $SM$ experiments, since the in-phase components in such experiments are background-free.

\begin{center}
\begin{table}[h]

\setlength{\tabcolsep}{4pt}
\begin{tabular}{cccccccc}
&&m=1&2&3&4&5&6\\
\hline
\multirow{2}{*}{q=1}&theo.&1&0&0&0&0&0\\
&exp.&1&0&0&0&0&0\\
\hline
\multirow{2}{*}{2}&theo.&+16/3$\pi$=0.849&0&-16/15$\pi$=-0.170&0&-16/105$\pi$=-0.024&0\\
&exp.&+0.844(27)&0&-0.168(5)&0&-0.023(1)&0\\
\hline
\multirow{2}{*}{3}&theo.&0&0&-1/3=-0.333&0&0&0\\
&exp.&0&0&-0.343(10)&0&0&0\\

\hline
\multirow{2}{*}{4}&theo.&-32/15$\pi$=-0.340&0&-32/21$\pi$=-0.243&0&+32/45$\pi$=+0.113&0\\
&exp.&-0.348(11)&0&-0.247(8)&0&+0.117(4)&0\\
\hline
\multirow{2}{*}{5}&theo.&0&0&0&0&+1/5=0.200&0\\
&exp.&0&0&0&0&+0.205(7)\\
\hline
\multirow{2}{*}{6}&theo.&+48/35$\pi$=+0.218&0&+16/27$\pi$=+0.094&0&+48/55$\pi$=+0.139&0\\
&exp.&+0.229(8)&0&+0.095(3)&0&+0.141(5)&0\\
\hline
\end{tabular}
\caption{Comparison of theoretical and experimental peak-peak amplitudes $d_{q,m}/d_{1,1}$ of the dispersive resonances in an AM experiment, normalized to the amplitude of the resonance with $q$=1 and $m$=1.
The row label $q$ refers to the frequency $q\,\omegaMod$ at which the signals are demodulated.
The column label $m$ denotes the position of the resonances at $\omega_L=\pm m\,\omegaMod$.\label{tab:comparison}}
\end{table}
\end{center}

\subsection{Hanle resonances}
As anticipated by the model, zero-field Hanle resonances are only observed on the in-phase spectra.
Here we do not attempt to analyze the relative magnitudes of the $\omega_L=0$ resonance(s) and the resonances with $m\neq 0$, since the widths and amplitudes of the zero-field level crossing resonances are strongly affected by  field inhomogeneities, i.e., by residual transverse field components.
In the experiments reported here, no effort was made to precisely cancel such transverse components, since the resonances of interest are only marginally affected by such field components.

\subsection{Alignment resonances}
The experimental quadrature spectra show small resonant structures at multiples of $\omega_L/2$ (Fig.~ \ref{fig:AMresonances}).
Our initial guess was that these resonances originate from an imperfect degree of circular polarization, i.e., from a small degree of linear polarization.
During completion of the present work we realized that these resonances are due to atomic alignment along the light propagation direction that is produced (and probed) by the circularly polarized laser beam.
It so happens that the production and detection of alignment is very inefficient on the $4\rightarrow 3$ transition that was chosen for the above case study.
Subsequent experiments, not shown here, have shown that these alignment contributions are much more pronounced on the other three hyperfine components of the $D_1$ line, and that the corresponding resonance amplitudes may even surpass those of the orientation-based signals in some cases.
Further studies of these resonances are underway in our laboratory.

\section{Summary and conclusion}
\label{sec:summary}
We have presented a quantitative algebraic model that describes the features of the complex magnetic resonance patterns that are observed in experiments using synchronous optical pumping with modulated circularly polarized light in a transverse magnetic field.
The model considers only contributions from atomic spin orientation (vector polarization), but is general in the sense that it can be applied---with a suitable choice of model parameters---to amplitude-, frequency-, and polarization-modulation experiments.
In all three types of experiments the polarization production and detection efficiencies are modulated in a periodic manner.
The model is also general in the sense that it applies to states with an arbitrary angular momentum $F$, since we used a description of the medium absorption in terms of irreducible multipole moments (here the $k$=1 vector polarization).

Explicit expressions are given for the signal background and the resonances that occur with DC (\emph{i.e.}, low-pass filtered) detection as well as with phase-sensitive detection of the signal components that oscillate in-phase and in quadrature with multiples of the modulation frequency.

As a case study we have recorded the in-phase and quadrature spectra with amplitude-modulated light, detected at the first six harmonics of the modulation frequency.
We find that the experimental results are well described by our model at a level of better than 5\%.

We have found indications for distinct signal contributions that arise from the production and detection of atomic spin alignment (tensor polarization) that are not included in the present model.
The theoretical modelling and experimental study of these alignment resonances is ongoing.

\section*{Note added}
While this paper was in review, we have tested our model predictions against polarization modulation experiments.
We find --- as in the amplitude modulation case reported above --- an excellent agreement (both for low-pass filtered and for lock-in detected signals) between experiments and model predictions.
Moreover, we have observed that the relative amplitudes of all resonances (in the $q$- and $m$- space) are perfectly well described by the model predictions, even when the power is increased to levels that exceed the the model's range of validity by a factor of 20!
These results will be published elsewhere.

\begin{acknowledgments}
The authors thank P.~Knowles, E.~Breschi and V.~Lebedev for critical reading of the manuscript and useful comments. This work was supported by the grant 200020\_140421/1 of the Swiss National Science Foundation  .

\end{acknowledgments}

\clearpage
\appendix
\section{Fourier structure of $P^\mathrm{TOE}(t)$}
\label{app:AppendixA}
The time dependence of the transmitted power $P^\mathrm{TOE}(t)$ is obtained by inserting expression \eqref{eq:ArbMod} for $\Gamma_p$
\begin{equation}
\Gamma_p(t)=\gamma_p\,\sum_{k=-\infty}^\infty g_{k}\,\cos\left(k\,\omegaMod t\right)
\equiv\gamma_p\,f_{\omegaMod}^\mathrm{TOE}(t)\,,
\label{eq:0}
\end{equation}
here with summation index $k$ instead of $m$, and the time dependent polarization \eqref{eq:SzArbMod}
\begin{widetext}
\begin{equation}
S_z^\mathrm{TOE}(t)=2\,G\sum_{m=-\infty}^\infty\,g_m\,\left[\mathcal{A}_{m}(\omega_L)\,\cos(m
\omegaMod t)
+\mathcal{D}_{m}(\omega_L)\,\sin(m\omegaMod t)\right]\,,
\label{eq:1}
\end{equation}
\end{widetext}
into the general system response function \eqref{eq:PTOE}
\begin{equation}
P^\mathrm{TOE}(t)=A^\mathrm{TOE}+B^\mathrm{TOE}\,f_{\omegaMod}^\mathrm{TOE}(t)+C^\mathrm{TOE}\,S_z^\mathrm{TOE}(t)
\,f_{\omegaMod}^\mathrm{TOE}(t)\,.
\label{eq:22}
\end{equation}
Expansion of the result yields
\begin{widetext}
\begin{align}
P^\mathrm{TOE}(t)&=
A^\mathrm{TOE}
+B^\mathrm{TOE}\,\sum_{k=-\infty}^\infty g_{k}\,\cos\left(k\,\omegaMod t\right)+\nonumber\\
&+2\,C^\mathrm{TOE}\,G\,\sum_{m=-\infty}^\infty g_{m}\,\mathcal{A}_{m}(\omega_L)\,\cos\left(m\,\omegaMod t\right)\,\sum_{k=-\infty}^\infty g_{k}\,\cos\left(k\,\omegaMod t\right)\nonumber\\
&+2\,C^\mathrm{TOE}\,G\,\sum_{m=-\infty}^\infty g_{m}\,\mathcal{D}_{m}(\omega_L)\,\sin\left(m\,\omegaMod t\right)\,\sum_{k=-\infty}^\infty g_{k}\,\cos\left(k\,\omegaMod t\right)\nonumber\\
&=A^\mathrm{TOE}
+B^\mathrm{TOE}\,\sum_{k=-\infty}^\infty g_{k}\,\cos\left(k\,\omegaMod t\right)+\nonumber\\
&+G\,C^\mathrm{TOE}\,\sum_{m=-\infty}^\infty \,g_m\,\mathcal{A}_{m}(\omega_L)
\,\sum_{k=-\infty}^\infty g_{k}\left[\,\cos\left[(k+m)\,\omegaMod t\right]+\cos\left((k-m)\,\omegaMod t\right)\right]\nonumber\\
&+G\,C^\mathrm{TOE}\,\sum_{m=-\infty}^\infty\,g_m\,\mathcal{D}_{m}(\omega_L)
\, \sum_{k=-\infty}^\infty g_{k}\left[\,\sin\left((k+m)\,\omegaMod t\right)-\sin\left((k-m)\,\omegaMod t\right)\right]
\label{eq:PofTfinal1}\,.
\end{align}
\end{widetext}%
After translating the summation indices $k\rightarrow q=k+m$ and $k\rightarrow q=k-m$ of the two terms in the sums of the $C^\mathrm{TOE}$ term, and renaming the summation index $k$ in the $B^\mathrm{TOE}$ term to $q$, the time dependent power can be rewritten as
\begin{widetext}
\begin{align}
P^\mathrm{TOE}(t)
&=A^\mathrm{TOE}
+B^\mathrm{TOE}\,\sum_{q=-\infty}^\infty g_{q}\,\cos\left(q\,\omegaMod t\right)+\nonumber\\
&\quad +G\,C^\mathrm{TOE}\,\sum_{m=-\infty}^\infty \,g_m\,\mathcal{A}_{m}(\omega_L)\,
\left[
\sum_{q=-\infty}^\infty g_{q-m}\cos\left(q\omegaMod t\right)
+\sum_{q=-\infty}^\infty g_{q+m}\cos\left(q\omegaMod t\right)
\right]\nonumber\\
&\quad+G\,C^\mathrm{TOE}\,\sum_{m=-\infty}^\infty \,g_m\,\mathcal{D}_{m}(\omega_L)\,
\left[
\sum_{q=-\infty}^\infty g_{q-m}\sin\left(q\omegaMod t\right)
+\sum_{q=-\infty}^\infty g_{q+m}\sin\left(q\omegaMod t\right)
\right]\nonumber\\
&=A^\mathrm{TOE}
+B^\mathrm{TOE}\,\sum_{q=-\infty}^\infty g_{q}\,\cos\left(q\,\omegaMod t\right)+\nonumber\\
&\quad +G\,C^\mathrm{TOE}\,\sum_{q=-\infty}^\infty \,
\sum_{m=-\infty}^\infty g_m\,(g_{q-m}+g_{q+m})\,\mathcal{A}_{m}(\omega_L)\cos\left(q\omegaMod t\right)\nonumber\\
&\quad +G\,C^\mathrm{TOE}\,\sum_{q=-\infty}^\infty \,
\sum_{m=-\infty}^\infty g_m\,(g_{q-m}-g_{q+m})\,\mathcal{D}_{m}(\omega_L)\sin\left(q\omegaMod t\right)\nonumber\\
&=A^\mathrm{TOE}+B^\mathrm{TOE}\,g_0
+\gamma_p\,C^\mathrm{TOE}\,\sum_{m=-\infty}^\infty g_m^2\,\mathcal{A}_{m}(\omega_L)\,+\nonumber\\
&\quad+B^\mathrm{TOE}\,\sum_{\genfrac{}{}{0pt}{}{q{=}-\infty}{q{\neq}0}}^\infty g_{q}\,\cos\left(q\,\omegaMod t\right)+\nonumber\\
&\quad +G\,C^\mathrm{TOE}\,\sum_{\genfrac{}{}{0pt}{}{q{=}-\infty}{q{\neq}0}}^\infty \,
\sum_{m=-\infty}^\infty g_m\,(g_{q-m}+g_{q+m})\,\mathcal{A}_{m}(\omega_L)\cos\left(q\omegaMod t\right)\nonumber\\
&\quad +G\,C^\mathrm{TOE}\,\sum_{\genfrac{}{}{0pt}{}{q{=}-\infty}{q{\neq}0}}^\infty \,
\sum_{m=-\infty}^\infty g_m\,(g_{q-m}-g_{q+m})\,\mathcal{D}_{m}(\omega_L)\sin\left(q\omegaMod t\right)
\label{eq:PofTfinal2}\,.
\end{align}
\end{widetext}%
In the last transformation we have extracted explicitly the time independent ($q=0$) terms from the sums over $q$.
Lock-in demodulation is done at frequencies $q\,\omega_L$, where $q$ is a positive non-zero integer.
As a last step, we therefore transform \eqref{eq:PofTfinal2} to have the sum over $q$ run over positive values only:
%
\begin{align}
P^\mathrm{TOE}(t)
&\equiv A^\mathrm{TOE}+B^\mathrm{TOE}\,g_0
+\,G\,C^\mathrm{TOE}\,\sum_{m=-\infty}^\infty g_m^2\,\mathcal{A}_{m}\,(\omega_L)+\nonumber\\
&\quad+2B^\mathrm{TOE}\,\sum_{q=1}^\infty g_{q}\,\cos\left(q\,\omegaMod t\right)+\nonumber\\
&\quad +G\,C^\mathrm{TOE}\,\sum_{q=1}^\infty \,
\sum_{m=-\infty}^\infty g_m\,(g_{q-m}+g_{q+m}+g_{-q-m}+g_{-q+m})\,\mathcal{A}_{m}(\omega_L)\cos\left(q\omegaMod t\right)\nonumber\\
&\quad +G\,C^\mathrm{TOE}\,\sum_{q=1}^\infty \,
\sum_{m=-\infty}^\infty g_m\,(g_{q-m}-g_{q+m}-(g_{-q-m}-g_{-q+m}))\,\mathcal{D}_{m}(\omega_L)\sin\left(q\omegaMod t\right)\nonumber\\
&=  A^\mathrm{TOE}+B^\mathrm{TOE}\,g_0
+\,G\,C^\mathrm{TOE}\,\sum_{m=-\infty}^\infty g_m^2\,\mathcal{A}_{m}\,(\omega_L)+\nonumber\\
&\quad+2B^\mathrm{TOE}\,\sum_{q=1}^\infty g_{q}\,\cos\left(q\,\omegaMod t\right)+\nonumber\\
&\quad +2\,G\,C^\mathrm{TOE}\,\sum_{q=1}^\infty \,
\sum_{m=-\infty}^\infty g_m\,(g_{q-m}+g_{q+m})\,\mathcal{A}_{m}(\omega_L)\cos\left(q\omegaMod t\right)\nonumber\\
&\quad +2\,G\,C^\mathrm{TOE}\,\sum_{q=1}^\infty \,
\sum_{m=-\infty}^\infty g_m\,(g_{q-m}-g_{q+m})\,\mathcal{D}_{m}(\omega_L)\sin\left(q\omegaMod t\right)\,,
\label{eq:PofTfinal-zg-1}
\end{align}
where we have used $g_{i}=g_{-i}$.
%


\subsection{Lock-in signals}  
The transmitted laser power contains a time independent term
\begin{equation}\label{eq:BGD}
    \mathcal{B}_\mathrm{DC}(\omega_L)\equiv A^\mathrm{TOE}+B^\mathrm{TOE}\,g_0
+\,G\,C^\mathrm{TOE}\,\sum_{m=-\infty}^\infty g_m^2\,\mathcal{A}_{m}\,(\omega_L)\,,
\end{equation}
%

in addition to the harmonic sum of all oscillating terms.

In the experiments, the time dependent terms that oscillate in phase and in quadrature at harmonics $q$ of the fundamental modulation frequency $\omegaMod$ are extracted by phase-sensitive (lock-in) detection.
We recall that lock-in extraction of the in-phase and quadrature amplitudes at the $q$-th harmonic (demodulation) consists in mixing (multiplying) the photodiode signal $P^\mathrm{TOE}(t)$ with $\cos(q\omegaMod t)$ and $\sin(q\omegaMod t)$, respectively,  followed by low-pass filtering of that product.
For calculational purposes, low-pass filtering is equivalent of taking the rms time average of the mixed signal, which is equivalent to replacing $\cos(q\omegaMod t)$ and $\sin(q\omegaMod t)$ by $1/\sqrt{2}$ in \eqref{eq:PofTfinal-zg-1} and setting to zero the time independent terms.
Applying this procedure to the signal \eqref{eq:PofTfinal-zg-1} we obtain the rms amplitudes of the in-phase signals following demodulation at $q\omegaMod$

%
%

%
\begin{widetext}
\begin{equation}
\frac{\mathcal{I}^\mathrm{TOE}_\mathrm{q}(\omega_L)}{\sqrt{2}}
=B^\mathrm{TOE}\,g_q
+G\,C^\mathrm{TOE}\,
\sum_{m=-\infty}^\infty g_m\,(g_{q-m}+g_{q+m})\,\mathcal{A}_{m}(\omega_L)
\label{eq:inPhaseAPP}
\end{equation}
\end{widetext}
and the corresponding quadrature amplitudes
\begin{widetext}
\begin{equation}
\frac{\mathcal{Q}^\mathrm{TOE}_\mathrm{q}(\omega_L)}{\sqrt{2}}
=G\,C^\mathrm{TOE}\,\sum_{m=-\infty}^\infty g_m\,(g_{q-m}-g_{q+m})\,\mathcal{D}_{m}(\omega_L)\,,
\label{eq:inQuadratureAPP}
\end{equation}
\end{widetext}
respectively.

The structure of the resulting spectra is discussed in the body of the paper.

\bibliography{bib.bib}

\begin{thebibliography}{22}%
\makeatletter
\providecommand \@ifxundefined [1]{%
 \@ifx{#1\undefined}
}%
\providecommand \@ifnum [1]{%
 \ifnum #1\expandafter \@firstoftwo
 \else \expandafter \@secondoftwo
 \fi
}%
\providecommand \@ifx [1]{%
 \ifx #1\expandafter \@firstoftwo
 \else \expandafter \@secondoftwo
 \fi
}%
\providecommand \natexlab [1]{#1}%
\providecommand \enquote  [1]{``#1''}%
\providecommand \bibnamefont  [1]{#1}%
\providecommand \bibfnamefont [1]{#1}%
\providecommand \citenamefont [1]{#1}%
\providecommand \href@noop [0]{\@secondoftwo}%
\providecommand \href [0]{\begingroup \@sanitize@url \@href}%
\providecommand \@href[1]{\@@startlink{#1}\@@href}%
\providecommand \@@href[1]{\endgroup#1\@@endlink}%
\providecommand \@sanitize@url [0]{\catcode `\\12\catcode `\$12\catcode
  `\&12\catcode `\#12\catcode `\^12\catcode `\_12\catcode `\%12\relax}%
\providecommand \@@startlink[1]{}%
\providecommand \@@endlink[0]{}%
\providecommand \url  [0]{\begingroup\@sanitize@url \@url }%
\providecommand \@url [1]{\endgroup\@href {#1}{\urlprefix }}%
\providecommand \urlprefix  [0]{URL }%
\providecommand \Eprint [0]{\href }%
\providecommand \doibase [0]{http://dx.doi.org/}%
\providecommand \selectlanguage [0]{\@gobble}%
\providecommand \bibinfo  [0]{\@secondoftwo}%
\providecommand \bibfield  [0]{\@secondoftwo}%
\providecommand \translation [1]{[#1]}%
\providecommand \BibitemOpen [0]{}%
\providecommand \bibitemStop [0]{}%
\providecommand \bibitemNoStop [0]{.\EOS\space}%
\providecommand \EOS [0]{\spacefactor3000\relax}%
\providecommand \BibitemShut  [1]{\csname bibitem#1\endcsname}%
\let\auto@bib@innerbib\@empty
\bibitem [{\citenamefont {Budker}\ \emph {et~al.}(2002)\citenamefont {Budker},
  \citenamefont {Gawlik}, \citenamefont {Kimball}, \citenamefont {Rochester},
  \citenamefont {Yashchuk},\ and\ \citenamefont
  {Weis}}]{BudkerWeis-RevModPhys.74.1153}%
  \BibitemOpen
  \bibfield  {author} {\bibinfo {author} {\bibfnamefont {D.}~\bibnamefont
  {Budker}}, \bibinfo {author} {\bibfnamefont {W.}~\bibnamefont {Gawlik}},
  \bibinfo {author} {\bibfnamefont {D.~F.}\ \bibnamefont {Kimball}}, \bibinfo
  {author} {\bibfnamefont {S.~M.}\ \bibnamefont {Rochester}}, \bibinfo {author}
  {\bibfnamefont {V.~V.}\ \bibnamefont {Yashchuk}}, \ and\ \bibinfo {author}
  {\bibfnamefont {A.}~\bibnamefont {Weis}},\ }\href {\doibase
  10.1103/RevModPhys.74.1153} {\bibfield  {journal} {\bibinfo  {journal} {Rev.
  Mod. Phys.}\ }\textbf {\bibinfo {volume} {74}},\ \bibinfo {pages} {1153}
  (\bibinfo {year} {2002})}\BibitemShut {NoStop}%
\bibitem [{\citenamefont {Budker}\ and\ \citenamefont
  {Romalis}(2007)}]{BudkerRomalisNature}%
  \BibitemOpen
  \bibfield  {author} {\bibinfo {author} {\bibfnamefont {D.}~\bibnamefont
  {Budker}}\ and\ \bibinfo {author} {\bibfnamefont {M.}~\bibnamefont
  {Romalis}},\ }\href {\doibase 10.1038/nphys566} {\bibfield  {journal}
  {\bibinfo  {journal} {Nat. Phys.}\ }\textbf {\bibinfo {volume} {3}},\
  \bibinfo {pages} {227} (\bibinfo {year} {2007})}\BibitemShut {NoStop}%
\bibitem [{\citenamefont {Groeger}\ \emph {et~al.}(2005)\citenamefont
  {Groeger}, \citenamefont {Pazgalev},\ and\ \citenamefont
  {Weis}}]{Groeger:LampvsLaser2004}%
  \BibitemOpen
  \bibfield  {author} {\bibinfo {author} {\bibfnamefont {S.}~\bibnamefont
  {Groeger}}, \bibinfo {author} {\bibfnamefont {A.}~\bibnamefont {Pazgalev}}, \
  and\ \bibinfo {author} {\bibfnamefont {A.}~\bibnamefont {Weis}},\ }\href
  {\doibase 10.1007/s00340-005-1773-x} {\bibfield  {journal} {\bibinfo
  {journal} {Appl. Phys. B}\ }\textbf {\bibinfo {volume} {80}},\ \bibinfo
  {pages} {645} (\bibinfo {year} {2005})}\BibitemShut {NoStop}%
\bibitem [{\citenamefont {Bison}\ \emph {et~al.}(2009)\citenamefont {Bison},
  \citenamefont {Castagna}, \citenamefont {Hofer}, \citenamefont {Knowles},
  \citenamefont {Schenker}, \citenamefont {Kasprzak}, \citenamefont {Saudan},\
  and\ \citenamefont {Weis}}]{Bison2009:APL19chSystem}%
  \BibitemOpen
  \bibfield  {author} {\bibinfo {author} {\bibfnamefont {G.}~\bibnamefont
  {Bison}}, \bibinfo {author} {\bibfnamefont {N.}~\bibnamefont {Castagna}},
  \bibinfo {author} {\bibfnamefont {A.}~\bibnamefont {Hofer}}, \bibinfo
  {author} {\bibfnamefont {P.}~\bibnamefont {Knowles}}, \bibinfo {author}
  {\bibfnamefont {J.~L.}\ \bibnamefont {Schenker}}, \bibinfo {author}
  {\bibfnamefont {M.}~\bibnamefont {Kasprzak}}, \bibinfo {author}
  {\bibfnamefont {H.}~\bibnamefont {Saudan}}, \ and\ \bibinfo {author}
  {\bibfnamefont {A.}~\bibnamefont {Weis}},\ }\href {\doibase
  10.1063/1.3255041} {\bibfield  {journal} {\bibinfo  {journal} {Appl. Phys.
  Lett.}\ }\textbf {\bibinfo {volume} {95}} (\bibinfo {year} {2009}),\
  10.1063/1.3255041}\BibitemShut {NoStop}%
\bibitem [{\citenamefont {Altarev}\ \emph {et~al.}(2009)\citenamefont
  {Altarev}, \citenamefont {Ban}, \citenamefont {Bison}, \citenamefont {Bodek},
  \citenamefont {Burghoff}, \citenamefont {Cvijovic}, \citenamefont {Daum},
  \citenamefont {Fierlinger}, \citenamefont {Gutsmiedl}, \citenamefont
  {Hampel}, \citenamefont {Heil}, \citenamefont {Henneck}, \citenamefont
  {Horras}, \citenamefont {Khomutov}, \citenamefont {Kirch}, \citenamefont
  {Kistryn}, \citenamefont {Knappe-Grüneberg}, \citenamefont {Knecht},
  \citenamefont {Knowles}, \citenamefont {Kozela}, \citenamefont {Kratz},
  \citenamefont {Kuchler}, \citenamefont {KuŸniak}, \citenamefont {Lauer},
  \citenamefont {Lauss}, \citenamefont {Lefort}, \citenamefont
  {Mtchedlishvili}, \citenamefont {Naviliat-Cuncic}, \citenamefont {Paul},
  \citenamefont {Pazgalev}, \citenamefont {Petzoldt}, \citenamefont {Pierre},
  \citenamefont {Plonka-Spehr}, \citenamefont {Quéméner}, \citenamefont
  {Rebreyend}, \citenamefont {Roccia}, \citenamefont {Rogel}, \citenamefont
  {Sander-Thoemmes}, \citenamefont {Schnabel}, \citenamefont {Severijns},
  \citenamefont {Sobolev}, \citenamefont {Stoepler}, \citenamefont {Trahms},
  \citenamefont {Weis}, \citenamefont {Wiehl}, \citenamefont {Zejma},\ and\
  \citenamefont {Zsigmond"}}]{PaulNIM}%
  \BibitemOpen
  \bibfield  {author} {\bibinfo {author} {\bibfnamefont {I.}~\bibnamefont
  {Altarev}}, \bibinfo {author} {\bibfnamefont {G.}~\bibnamefont {Ban}},
  \bibinfo {author} {\bibfnamefont {G.}~\bibnamefont {Bison}}, \bibinfo
  {author} {\bibfnamefont {K.}~\bibnamefont {Bodek}}, \bibinfo {author}
  {\bibfnamefont {M.}~\bibnamefont {Burghoff}}, \bibinfo {author}
  {\bibfnamefont {M.}~\bibnamefont {Cvijovic}}, \bibinfo {author}
  {\bibfnamefont {M.}~\bibnamefont {Daum}}, \bibinfo {author} {\bibfnamefont
  {P.}~\bibnamefont {Fierlinger}}, \bibinfo {author} {\bibfnamefont
  {E.}~\bibnamefont {Gutsmiedl}}, \bibinfo {author} {\bibfnamefont
  {G.}~\bibnamefont {Hampel}}, \bibinfo {author} {\bibfnamefont
  {W.}~\bibnamefont {Heil}}, \bibinfo {author} {\bibfnamefont {R.}~\bibnamefont
  {Henneck}}, \bibinfo {author} {\bibfnamefont {M.}~\bibnamefont {Horras}},
  \bibinfo {author} {\bibfnamefont {N.}~\bibnamefont {Khomutov}}, \bibinfo
  {author} {\bibfnamefont {K.}~\bibnamefont {Kirch}}, \bibinfo {author}
  {\bibfnamefont {S.}~\bibnamefont {Kistryn}}, \bibinfo {author} {\bibfnamefont
  {S.}~\bibnamefont {Knappe-Grüneberg}}, \bibinfo {author} {\bibfnamefont
  {A.}~\bibnamefont {Knecht}}, \bibinfo {author} {\bibfnamefont
  {P.}~\bibnamefont {Knowles}}, \bibinfo {author} {\bibfnamefont
  {A.}~\bibnamefont {Kozela}}, \bibinfo {author} {\bibfnamefont
  {J.}~\bibnamefont {Kratz}}, \bibinfo {author} {\bibfnamefont
  {F.}~\bibnamefont {Kuchler}}, \bibinfo {author} {\bibfnamefont
  {M.}~\bibnamefont {KuŸniak}}, \bibinfo {author} {\bibfnamefont
  {T.}~\bibnamefont {Lauer}}, \bibinfo {author} {\bibfnamefont
  {B.}~\bibnamefont {Lauss}}, \bibinfo {author} {\bibfnamefont
  {T.}~\bibnamefont {Lefort}}, \bibinfo {author} {\bibfnamefont
  {A.}~\bibnamefont {Mtchedlishvili}}, \bibinfo {author} {\bibfnamefont
  {O.}~\bibnamefont {Naviliat-Cuncic}}, \bibinfo {author} {\bibfnamefont
  {S.}~\bibnamefont {Paul}}, \bibinfo {author} {\bibfnamefont {A.}~\bibnamefont
  {Pazgalev}}, \bibinfo {author} {\bibfnamefont {G.}~\bibnamefont {Petzoldt}},
  \bibinfo {author} {\bibfnamefont {E.}~\bibnamefont {Pierre}}, \bibinfo
  {author} {\bibfnamefont {C.}~\bibnamefont {Plonka-Spehr}}, \bibinfo {author}
  {\bibfnamefont {G.}~\bibnamefont {Quéméner}}, \bibinfo {author}
  {\bibfnamefont {D.}~\bibnamefont {Rebreyend}}, \bibinfo {author}
  {\bibfnamefont {S.}~\bibnamefont {Roccia}}, \bibinfo {author} {\bibfnamefont
  {G.}~\bibnamefont {Rogel}}, \bibinfo {author} {\bibfnamefont
  {T.}~\bibnamefont {Sander-Thoemmes}}, \bibinfo {author} {\bibfnamefont
  {A.}~\bibnamefont {Schnabel}}, \bibinfo {author} {\bibfnamefont
  {N.}~\bibnamefont {Severijns}}, \bibinfo {author} {\bibfnamefont
  {Y.}~\bibnamefont {Sobolev}}, \bibinfo {author} {\bibfnamefont
  {R.}~\bibnamefont {Stoepler}}, \bibinfo {author} {\bibfnamefont
  {L.}~\bibnamefont {Trahms}}, \bibinfo {author} {\bibfnamefont
  {A.}~\bibnamefont {Weis}}, \bibinfo {author} {\bibfnamefont {N.}~\bibnamefont
  {Wiehl}}, \bibinfo {author} {\bibfnamefont {J.}~\bibnamefont {Zejma}}, \ and\
  \bibinfo {author} {\bibfnamefont {G.}~\bibnamefont {Zsigmond"}},\ }\href
  {\doibase 10.1016/j.nima.2009.07.046} {\bibfield  {journal} {\bibinfo
  {journal} {Nucl. Instr. Meth. A}\ }\textbf {\bibinfo {volume} {611}},\
  \bibinfo {pages} {133 } (\bibinfo {year} {2009})}\BibitemShut {NoStop}%
\bibitem [{\citenamefont {Alexandrov}\ \emph {et~al.}(1996)\citenamefont
  {Alexandrov}, \citenamefont {Balabas}, \citenamefont {Pasgalev},
  \citenamefont {Vershovskii},\ and\ \citenamefont {Yakobson}}]{AlexandrovMx}%
  \BibitemOpen
  \bibfield  {author} {\bibinfo {author} {\bibfnamefont {E.}~\bibnamefont
  {Alexandrov}}, \bibinfo {author} {\bibfnamefont {M.}~\bibnamefont {Balabas}},
  \bibinfo {author} {\bibfnamefont {A.}~\bibnamefont {Pasgalev}}, \bibinfo
  {author} {\bibfnamefont {A.}~\bibnamefont {Vershovskii}}, \ and\ \bibinfo
  {author} {\bibfnamefont {N.}~\bibnamefont {Yakobson}},\ }\href@noop {}
  {\bibfield  {journal} {\bibinfo  {journal} {Laser Phys.}\ }\textbf {\bibinfo
  {volume} {6}},\ \bibinfo {pages} {244} (\bibinfo {year} {1996})}\BibitemShut
  {NoStop}%
\bibitem [{\citenamefont {Groeger}\ \emph {et~al.}(2006)\citenamefont
  {Groeger}, \citenamefont {Bison}, \citenamefont {Knowles}, \citenamefont
  {Wynands},\ and\ \citenamefont {Weis}}]{GroegerMx-2007}%
  \BibitemOpen
  \bibfield  {author} {\bibinfo {author} {\bibfnamefont {S.}~\bibnamefont
  {Groeger}}, \bibinfo {author} {\bibfnamefont {G.}~\bibnamefont {Bison}},
  \bibinfo {author} {\bibfnamefont {P.}~\bibnamefont {Knowles}}, \bibinfo
  {author} {\bibfnamefont {R.}~\bibnamefont {Wynands}}, \ and\ \bibinfo
  {author} {\bibfnamefont {A.}~\bibnamefont {Weis}},\ }\href {\doibase
  10.1016/j.sna.2005.09.036} {\bibfield  {journal} {\bibinfo  {journal}
  {Sensors and Actuators A-Physical}\ }\textbf {\bibinfo {volume} {129}},\
  \bibinfo {pages} {1 } (\bibinfo {year} {2006})}\BibitemShut {NoStop}%
\bibitem [{\citenamefont {Acosta}\ \emph {et~al.}(2006)\citenamefont {Acosta},
  \citenamefont {Ledbetter}, \citenamefont {Rochester}, \citenamefont {Budker},
  \citenamefont {Jackson~Kimball}, \citenamefont {Hovde}, \citenamefont
  {Gawlik}, \citenamefont {Pustelny}, \citenamefont {Zachorowski},\ and\
  \citenamefont {Yashchuk}}]{Budker:BalRot.PhysRevA.73.053404}%
  \BibitemOpen
  \bibfield  {author} {\bibinfo {author} {\bibfnamefont {V.}~\bibnamefont
  {Acosta}}, \bibinfo {author} {\bibfnamefont {M.~P.}\ \bibnamefont
  {Ledbetter}}, \bibinfo {author} {\bibfnamefont {S.~M.}\ \bibnamefont
  {Rochester}}, \bibinfo {author} {\bibfnamefont {D.}~\bibnamefont {Budker}},
  \bibinfo {author} {\bibfnamefont {D.~F.}\ \bibnamefont {Jackson~Kimball}},
  \bibinfo {author} {\bibfnamefont {D.~C.}\ \bibnamefont {Hovde}}, \bibinfo
  {author} {\bibfnamefont {W.}~\bibnamefont {Gawlik}}, \bibinfo {author}
  {\bibfnamefont {S.}~\bibnamefont {Pustelny}}, \bibinfo {author}
  {\bibfnamefont {J.}~\bibnamefont {Zachorowski}}, \ and\ \bibinfo {author}
  {\bibfnamefont {V.~V.}\ \bibnamefont {Yashchuk}},\ }\href {\doibase
  10.1103/PhysRevA.73.053404} {\bibfield  {journal} {\bibinfo  {journal} {Phys.
  Rev. A}\ }\textbf {\bibinfo {volume} {73}},\ \bibinfo {pages} {053404}
  (\bibinfo {year} {2006})}\BibitemShut {NoStop}%
\bibitem [{\citenamefont {Gawlik}\ \emph {et~al.}(2006)\citenamefont {Gawlik},
  \citenamefont {Krzemien}, \citenamefont {Pustelny}, \citenamefont {Sangla},
  \citenamefont {Zachorowski}, \citenamefont {Graf}, \citenamefont {Sushkov},\
  and\ \citenamefont {Budker}}]{GawlikAM}%
  \BibitemOpen
  \bibfield  {author} {\bibinfo {author} {\bibfnamefont {W.}~\bibnamefont
  {Gawlik}}, \bibinfo {author} {\bibfnamefont {L.}~\bibnamefont {Krzemien}},
  \bibinfo {author} {\bibfnamefont {S.}~\bibnamefont {Pustelny}}, \bibinfo
  {author} {\bibfnamefont {D.}~\bibnamefont {Sangla}}, \bibinfo {author}
  {\bibfnamefont {J.}~\bibnamefont {Zachorowski}}, \bibinfo {author}
  {\bibfnamefont {M.}~\bibnamefont {Graf}}, \bibinfo {author} {\bibfnamefont
  {A.}~\bibnamefont {Sushkov}}, \ and\ \bibinfo {author} {\bibfnamefont
  {D.}~\bibnamefont {Budker}},\ }\href {\doibase 10.1063/1.2190457} {\bibfield
  {journal} {\bibinfo  {journal} {Appl. Phys. Lett.}\ }\textbf {\bibinfo
  {volume} {88}} (\bibinfo {year} {2006}),\ 10.1063/1.2190457}\BibitemShut
  {NoStop}%
\bibitem [{\citenamefont {Schultze}\ \emph {et~al.}(2012)\citenamefont
  {Schultze}, \citenamefont {IJsselsteijn}, \citenamefont {Scholtes},
  \citenamefont {Woetzel},\ and\ \citenamefont {Meyer}}]{JENA:Schultze:12}%
  \BibitemOpen
  \bibfield  {author} {\bibinfo {author} {\bibfnamefont {V.}~\bibnamefont
  {Schultze}}, \bibinfo {author} {\bibfnamefont {R.}~\bibnamefont
  {IJsselsteijn}}, \bibinfo {author} {\bibfnamefont {T.}~\bibnamefont
  {Scholtes}}, \bibinfo {author} {\bibfnamefont {S.}~\bibnamefont {Woetzel}}, \
  and\ \bibinfo {author} {\bibfnamefont {H.-G.}\ \bibnamefont {Meyer}},\ }\href
  {\doibase 10.1364/OE.20.014201} {\bibfield  {journal} {\bibinfo  {journal}
  {Opt. Expr.}\ }\textbf {\bibinfo {volume} {20}},\ \bibinfo {pages} {14201}
  (\bibinfo {year} {2012})}\BibitemShut {NoStop}%
\bibitem [{\citenamefont {Bell}\ and\ \citenamefont
  {Bloom}(1961{\natexlab{a}})}]{Bell:1961:ODS:PhysRevLett.6.280}%
  \BibitemOpen
  \bibfield  {author} {\bibinfo {author} {\bibfnamefont {W.~E.}\ \bibnamefont
  {Bell}}\ and\ \bibinfo {author} {\bibfnamefont {A.~L.}\ \bibnamefont
  {Bloom}},\ }\href {\doibase 10.1103/PhysRevLett.6.280} {\bibfield  {journal}
  {\bibinfo  {journal} {Phys. Rev. Lett.}\ }\textbf {\bibinfo {volume} {6}},\
  \bibinfo {pages} {280} (\bibinfo {year} {1961}{\natexlab{a}})}\BibitemShut
  {NoStop}%
\bibitem [{\citenamefont {Bell}\ and\ \citenamefont
  {Bloom}(1961{\natexlab{b}})}]{BellBloom:Forbidden:PhysRevLett.6.623}%
  \BibitemOpen
  \bibfield  {author} {\bibinfo {author} {\bibfnamefont {W.~E.}\ \bibnamefont
  {Bell}}\ and\ \bibinfo {author} {\bibfnamefont {A.~L.}\ \bibnamefont
  {Bloom}},\ }\href {\doibase 10.1103/PhysRevLett.6.623} {\bibfield  {journal}
  {\bibinfo  {journal} {Phys. Rev. Lett.}\ }\textbf {\bibinfo {volume} {6}},\
  \bibinfo {pages} {623} (\bibinfo {year} {1961}{\natexlab{b}})}\BibitemShut
  {NoStop}%
\bibitem [{\citenamefont {Aleksandrov}(1973)}]{N:Alixandrov:1973}%
  \BibitemOpen
  \bibfield  {author} {\bibinfo {author} {\bibfnamefont {E.~B.}\ \bibnamefont
  {Aleksandrov}},\ }\href@noop {} {\bibfield  {journal} {\bibinfo  {journal}
  {Soviet Physics Uspekhi}\ }\textbf {\bibinfo {volume} {15}},\ \bibinfo
  {pages} {436} (\bibinfo {year} {1973})}\BibitemShut {NoStop}%
\bibitem [{\citenamefont {Alexandrov}\ \emph {et~al.}(2005)\citenamefont
  {Alexandrov}, \citenamefont {Auzinsh}, \citenamefont {Budker}, \citenamefont
  {Kimball}, \citenamefont {Rochester},\ and\ \citenamefont
  {Yashchuk}}]{N:Alexandrov-Budker:2005}%
  \BibitemOpen
  \bibfield  {author} {\bibinfo {author} {\bibfnamefont {E.~B.}\ \bibnamefont
  {Alexandrov}}, \bibinfo {author} {\bibfnamefont {M.}~\bibnamefont {Auzinsh}},
  \bibinfo {author} {\bibfnamefont {D.}~\bibnamefont {Budker}}, \bibinfo
  {author} {\bibfnamefont {D.~F.}\ \bibnamefont {Kimball}}, \bibinfo {author}
  {\bibfnamefont {S.~M.}\ \bibnamefont {Rochester}}, \ and\ \bibinfo {author}
  {\bibfnamefont {V.~V.}\ \bibnamefont {Yashchuk}},\ }\href@noop {} {\bibfield
  {journal} {\bibinfo  {journal} {J. Opt. Soc. Am. B}\ }\textbf {\bibinfo
  {volume} {22}},\ \bibinfo {pages} {7} (\bibinfo {year} {2005})}\BibitemShut
  {NoStop}%
\bibitem [{\citenamefont {Ben-Kish}\ and\ \citenamefont
  {Romalis}(2010)}]{N:Romalis:2010}%
  \BibitemOpen
  \bibfield  {author} {\bibinfo {author} {\bibfnamefont {A.}~\bibnamefont
  {Ben-Kish}}\ and\ \bibinfo {author} {\bibfnamefont {M.~V.}\ \bibnamefont
  {Romalis}},\ }\href@noop {} {\bibfield  {journal} {\bibinfo  {journal} {Phys.
  Rev. Lett.}\ }\textbf {\bibinfo {volume} {105}},\ \bibinfo {pages} {193601}
  (\bibinfo {year} {2010})}\BibitemShut {NoStop}%
\bibitem [{\citenamefont {Malakyan}\ \emph {et~al.}(2004)\citenamefont
  {Malakyan}, \citenamefont {Rochester}, \citenamefont {Budker}, \citenamefont
  {Kimball},\ and\ \citenamefont {Yashchuk}}]{Malakyan:PhysRevA.69.013817}%
  \BibitemOpen
  \bibfield  {author} {\bibinfo {author} {\bibfnamefont {Y.~P.}\ \bibnamefont
  {Malakyan}}, \bibinfo {author} {\bibfnamefont {S.~M.}\ \bibnamefont
  {Rochester}}, \bibinfo {author} {\bibfnamefont {D.}~\bibnamefont {Budker}},
  \bibinfo {author} {\bibfnamefont {D.~F.}\ \bibnamefont {Kimball}}, \ and\
  \bibinfo {author} {\bibfnamefont {V.~V.}\ \bibnamefont {Yashchuk}},\ }\href
  {\doibase 10.1103/PhysRevA.69.013817} {\bibfield  {journal} {\bibinfo
  {journal} {Phys. Rev. A}\ }\textbf {\bibinfo {volume} {69}},\ \bibinfo
  {pages} {013817} (\bibinfo {year} {2004})}\BibitemShut {NoStop}%
\bibitem [{\citenamefont {Nagel}\ \emph {et~al.}(1998)\citenamefont {Nagel},
  \citenamefont {Graf}, \citenamefont {Naumov}, \citenamefont {Mariotti},
  \citenamefont {Biancalana}, \citenamefont {Meschede},\ and\ \citenamefont
  {Wynands}}]{Nagel:DarkStateMagnetometers1998}%
  \BibitemOpen
  \bibfield  {author} {\bibinfo {author} {\bibfnamefont {A.}~\bibnamefont
  {Nagel}}, \bibinfo {author} {\bibfnamefont {L.}~\bibnamefont {Graf}},
  \bibinfo {author} {\bibfnamefont {A.}~\bibnamefont {Naumov}}, \bibinfo
  {author} {\bibfnamefont {E.}~\bibnamefont {Mariotti}}, \bibinfo {author}
  {\bibfnamefont {V.}~\bibnamefont {Biancalana}}, \bibinfo {author}
  {\bibfnamefont {D.}~\bibnamefont {Meschede}}, \ and\ \bibinfo {author}
  {\bibfnamefont {R.}~\bibnamefont {Wynands}},\ }\href
  {http://stacks.iop.org/0295-5075/44/i=1/a=031} {\bibfield  {journal}
  {\bibinfo  {journal} {Europh. Lett.}\ }\textbf {\bibinfo {volume} {44}},\
  \bibinfo {pages} {31} (\bibinfo {year} {1998})}\BibitemShut {NoStop}%
\bibitem [{\citenamefont {Wynands}\ \emph {et~al.}(1998)\citenamefont
  {Wynands}, \citenamefont {Nagel}, \citenamefont {Brandt}, \citenamefont
  {Meschede},\ and\ \citenamefont {Weis}}]{NagelWynandsWeis:PhysRevA.58.196}%
  \BibitemOpen
  \bibfield  {author} {\bibinfo {author} {\bibfnamefont {R.}~\bibnamefont
  {Wynands}}, \bibinfo {author} {\bibfnamefont {A.}~\bibnamefont {Nagel}},
  \bibinfo {author} {\bibfnamefont {S.}~\bibnamefont {Brandt}}, \bibinfo
  {author} {\bibfnamefont {D.}~\bibnamefont {Meschede}}, \ and\ \bibinfo
  {author} {\bibfnamefont {A.}~\bibnamefont {Weis}},\ }\href {\doibase
  10.1103/PhysRevA.58.196} {\bibfield  {journal} {\bibinfo  {journal} {Phys.
  Rev. A}\ }\textbf {\bibinfo {volume} {58}},\ \bibinfo {pages} {196} (\bibinfo
  {year} {1998})}\BibitemShut {NoStop}%
\bibitem [{\citenamefont {Alzetta}\ \emph {et~al.}(1976)\citenamefont
  {Alzetta}, \citenamefont {Gozzini}, \citenamefont {Moi},\ and\ \citenamefont
  {Orriols}}]{Alzetta:Visualization:1976}%
  \BibitemOpen
  \bibfield  {author} {\bibinfo {author} {\bibfnamefont {G.}~\bibnamefont
  {Alzetta}}, \bibinfo {author} {\bibfnamefont {A.}~\bibnamefont {Gozzini}},
  \bibinfo {author} {\bibfnamefont {L.}~\bibnamefont {Moi}}, \ and\ \bibinfo
  {author} {\bibfnamefont {G.}~\bibnamefont {Orriols}},\ }\href {\doibase
  10.1007/BF02749417} {\bibfield  {journal} {\bibinfo  {journal} {Nuovo Cimento
  B}\ }\textbf {\bibinfo {volume} {36}},\ \bibinfo {pages} {5} (\bibinfo {year}
  {1976})}\BibitemShut {NoStop}%
\bibitem [{\citenamefont {Wolfram~Research}(2010)}]{Mathematica}%
  \BibitemOpen
  \bibfield  {author} {\bibinfo {author} {\bibfnamefont {I.}~\bibnamefont
  {Wolfram~Research}},\ }\href@noop {} {\emph {\bibinfo {title}
  {Mathematica}}},\ \bibinfo {edition} {version 8.0}\ ed.\ (\bibinfo
  {publisher} {Wolfram Research, Inc.},\ \bibinfo {address} {Champaign,
  Illinois},\ \bibinfo {year} {2010})\BibitemShut {NoStop}%
\bibitem [{\citenamefont {Castagna}\ and\ \citenamefont
  {Weis}(2011)}]{Castagna:PhysRevA.84.053421}%
  \BibitemOpen
  \bibfield  {author} {\bibinfo {author} {\bibfnamefont {N.}~\bibnamefont
  {Castagna}}\ and\ \bibinfo {author} {\bibfnamefont {A.}~\bibnamefont
  {Weis}},\ }\href {\doibase 10.1103/PhysRevA.84.053421} {\bibfield  {journal}
  {\bibinfo  {journal} {Phys. Rev. A}\ }\textbf {\bibinfo {volume} {84}},\
  \bibinfo {pages} {053421} (\bibinfo {year} {2011})}\BibitemShut {NoStop}%
\bibitem [{\citenamefont {Castagna}\ \emph {et~al.}(2009)\citenamefont
  {Castagna}, \citenamefont {Bison}, \citenamefont {Domenico}, \citenamefont
  {Hofer}, \citenamefont {Knowles}, \citenamefont {Macchione}, \citenamefont
  {Saudan},\ and\ \citenamefont {Weis}}]{castagna:Cells:2009}%
  \BibitemOpen
  \bibfield  {author} {\bibinfo {author} {\bibfnamefont {N.}~\bibnamefont
  {Castagna}}, \bibinfo {author} {\bibfnamefont {G.}~\bibnamefont {Bison}},
  \bibinfo {author} {\bibfnamefont {G.}~\bibnamefont {Domenico}}, \bibinfo
  {author} {\bibfnamefont {A.}~\bibnamefont {Hofer}}, \bibinfo {author}
  {\bibfnamefont {P.}~\bibnamefont {Knowles}}, \bibinfo {author} {\bibfnamefont
  {C.}~\bibnamefont {Macchione}}, \bibinfo {author} {\bibfnamefont
  {H.}~\bibnamefont {Saudan}}, \ and\ \bibinfo {author} {\bibfnamefont
  {A.}~\bibnamefont {Weis}},\ }\href {\doibase 10.1007/s00340-009-3464-5}
  {\bibfield  {journal} {\bibinfo  {journal} {Appl. Phys. B}\ }\textbf
  {\bibinfo {volume} {96}},\ \bibinfo {pages} {763} (\bibinfo {year}
  {2009})}\BibitemShut {NoStop}%
\end{thebibliography}%

\end{document}